\documentclass[aps,pra,floatfix,10pt,twocolumn,superscriptaddress,showpacs,longbibliography]{revtex4-1}

\usepackage{graphicx} 
\usepackage{amssymb}
\usepackage{amsmath}
\usepackage{color} 
\usepackage{bm}
\usepackage[english]{babel}
\usepackage{hyperref}

\newcommand {\ket}[1] {|#1 \rangle} \newcommand {\bra}[1] {\langle#1 |}

\newcommand {\ds}{\downarrow}
\newcommand {\us}{\uparrow}

\newcommand{\bal}{\boldsymbol \alpha}
\newcommand{\bbe}{\boldsymbol \beta}

\begin{document}

\title{Entanglement Growth in Quench Dynamics With Variable Range Interactions}

\author{J.~Schachenmayer} \affiliation{Department of Physics and
Astronomy, University of Pittsburgh, Pittsburgh, Pennsylvania 15260, USA}
\author{B.~P.~Lanyon} \author{C.~F.~Roos} \affiliation{Institute for Quantum Optics and Quantum Information of the Austrian Academy of Sciences and Institute for Experimental Physics,
University of Innsbruck, Innsbruck, Austria}

\author{A.~J.~Daley} \affiliation{Department of Physics and Astronomy,
University of Pittsburgh, Pittsburgh, Pennsylvania 15260, USA}

\date{\today}

\begin{abstract} 

Studying entanglement growth in quantum dynamics provides both insight into the underlying microscopic processes and information about the complexity of the quantum states, which is related to the efficiency of simulations on classical computers. Recently, experiments with trapped ions, polar molecules, and Rydberg excitations have provided new opportunities to observe dynamics with long-range interactions. We explore nonequilibrium coherent dynamics after a quantum quench in such systems, identifying qualitatively different behavior as the exponent of algebraically decaying spin-spin interactions in a transverse Ising chain is varied. Computing the build-up of bipartite entanglement as well as mutual information between distant spins, we identify linear growth of entanglement entropy corresponding to propagation of quasiparticles for shorter range interactions, with the maximum rate of growth occurring when the Hamiltonian parameters match those for the quantum phase transition. Counter-intuitively, the growth of bipartite entanglement for long-range interactions is only logarithmic for most regimes, i.e., substantially slower than for shorter range interactions. Experiments with trapped ions allow for the realization of this system with a tunable interaction range, and we show that the different phenomena are robust for finite system sizes and in the presence of noise. These results can act as a direct guide for the generation of large-scale entanglement in such experiments, towards a regime where the entanglement growth can render existing classical simulations inefficient.

\end{abstract}

\maketitle

\section{Introduction}

Advances with atomic molecular and optical (AMO) systems, including cold atoms, entangled photons, and trapped ions, have rapidly opened possibilities to explore many-body physics in a highly controllable way \cite{bloch_quantum_2012,blatt_quantum_2012,aspuru-guzik_photonic_2012}. A key example of this is the new possibility to explore coherent nonequilibrium dynamics in a closed many-body system, e.g., the dynamics induced by quantum quenches \cite{rigol_hard-core_2006,cazalilla_effect_2006,calabrese_time_2006, kollath_quench_2007, lauchli_spreading_2008,PhysRevA.79.060302,worm_relaxation_2012,foss-feig_nonequilibrium_2013, hazzard_far--equilibrium_2013,PhysRevA.79.060302, PhysRevB.81.100412}. There have been several recent quench experiments with cold atoms in optical lattices, which not only probe the microscopic behavior of the system, e.g., the propagation of quasiparticles \cite{cheneau_light-cone-like_2012}, but also indicate the possibility to probe dynamics beyond the regimes that are currently accessible to classical simulations \cite{trotzky_probing_2012,schuch_entropy_2008,verstraete_matrix_2006}. In this context, the growth of entanglement in the system underlies the complexity of simulating the dynamics classically. Demonstration of large entanglement growth after a quantum quench would be a crucial step in demonstrating the possibility to use these systems as controllable quantum simulators, effectively using experimental systems to compute dynamics in a way that exceeds the capabilities of classical computations \cite{cirac_goals_2012}. 

Systems of trapped ions are a very promising candidate for realizing a quantum simulator, because of the control already demonstrated in the development of gate-based quantum computation and simulation \cite{haffner_quantum_2008,wineland_quantum_2011,lanyon_universal_2011} 
with these systems, and the ability to make measurements by state tomography \cite{roos_bell_2004}. Recently, analogue quantum simulation of interacting spin systems \cite{porras_effective_2004,schneider_experimental_2012} was also realized in ion traps \cite{friedenauer_simulating_2008,islam_onset_2011,kim_quantum_2011}, with a key novel element being the possibility to realize variable-range interactions \cite{Britton:2012, islam_emergence_2013,richerme_trapped-ion_2013,PhysRevLett.103.120502, 1367-2630-14-9-095024, PhysRevA.87.013422}, as shown in Fig.~1, in contrast to the short-range interactions of neutral atoms, or the dipole-dipole interactions possible with polar molecules.

\begin{figure}[tb] 
\begin{center}
\includegraphics[width=0.45\textwidth]{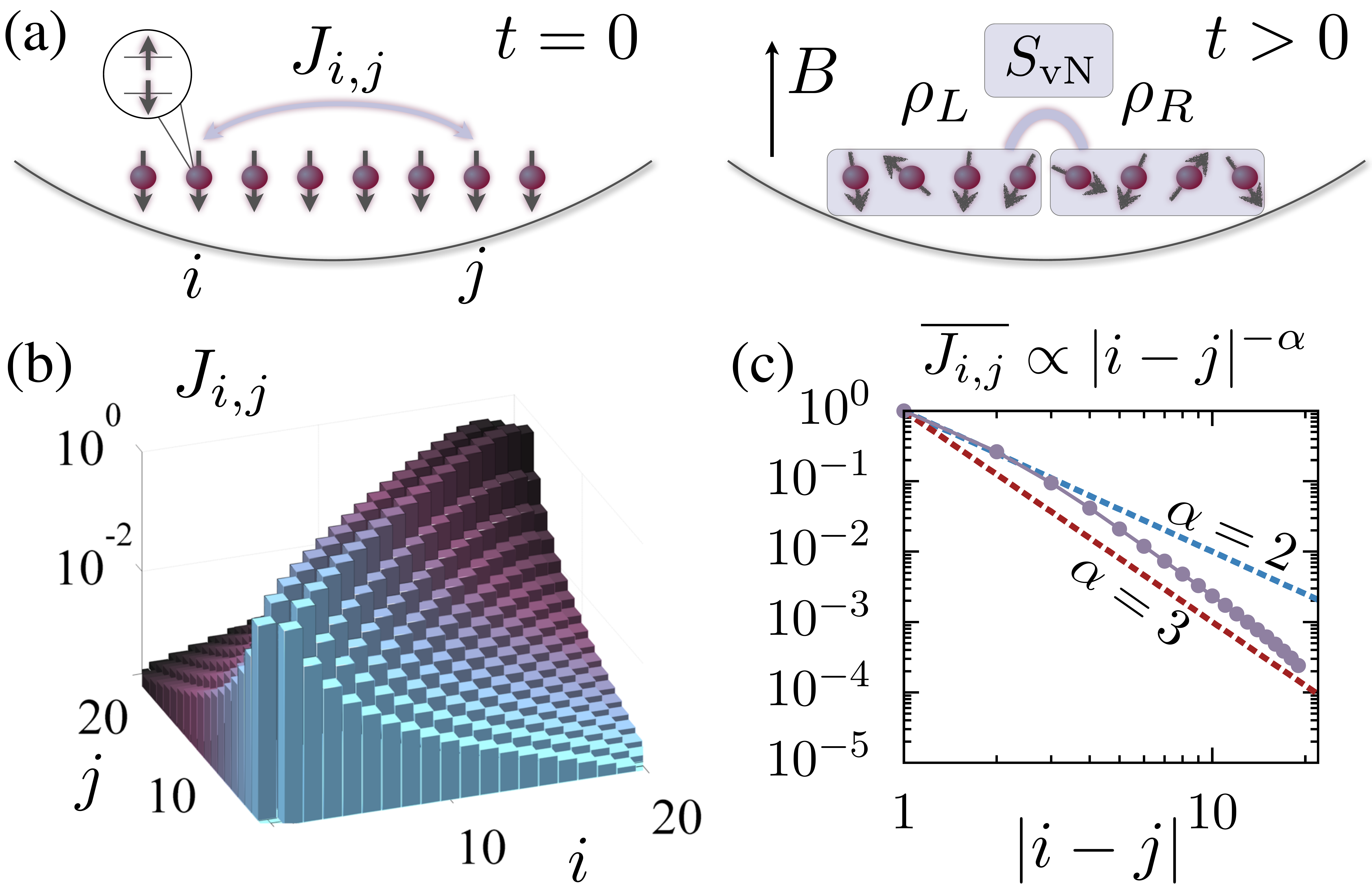} 
\end{center}
\caption{(a) Illustration of the quench experiment. We consider a linear chain of ions (effective spin-model) with long-range interactions. Initially, all spins are fully polarized along the axis of the magnetic field $B$. After a time-evolution spatial entanglement entropy ($S_{\rm vN}$) builds up between blocks of the system. (b) A typical calculated experimental interaction matrix for $20$ ions (see text for further details and parameters).  (c) The decay of the interactions with a tunable decay exponent $\alpha$. Here, the grey dots show the mean interactions from panel (b).
\label{fig:fig1}}
\end{figure}

So far, these variable-range interactions were discussed primarily in the case of ground-state calculations and near-adiabatic dynamics. Here, we explore nonequilibrium coherent dynamics after a quantum quench in these systems, identifying qualitatively different behavior as the exponent $\alpha$ of algebraically decaying spin-spin interactions is varied. Beginning with all spins aligned with a transverse field, we use a combination of analytical and numerical methods to compute the dynamics after the Ising interactions are quenched on, incorporating matrix product operator techniques  \cite{crosswhite_applying_2008,pirvu_matrix_2010,frowis_tensor_2010,vidal_efficient_2004,
  white_real-time_2004, daley_time-dependent_2004, schollwock_density-matrix_2011,verstraete_matrix_2008} to treat variable long-range interactions with up to 50 spins. 
  
In particular, we investigate the build up of bipartite entanglement in the chain as well as mutual information between distant spins  \cite{calabrese_time_2006, kollath_quench_2007, lauchli_spreading_2008,PhysRevA.79.060302}. For interactions with $\alpha\gtrsim 1$, we show that the behavior is qualitatively similar to nearest-neighbor interactions, with correlation build up well described by the propagation of quasiparticles at a rate equal to or slower than the Lieb-Robinson bound \cite{lieb_finite_1972, calabrese_evolution_2005, cramer_locality_2008}. This leads to a linear increase in bipartite entanglement in time, so that the dynamics cannot be efficiently computed in existing classical simulations beyond short times \cite{verstraete_matrix_2006,schuch_entropy_2008}. Interestingly, in this limit we find that the maximum growth rate of bipartite entanglement, even in small systems, occurs when we quench the interaction strength to the value corresponding to the quantum phase transition point, shifting accordingly for varying $\alpha$. 

For interactions with $\alpha \lesssim 1$, we observe qualitatively different behavior. Counterintuitively, quenches above the critical point for these long-range interactions lead only to a logarithmic increase of bipartite entanglement in time, so that in this regime, long-range interactions produce a slower growth of entanglement than short-range interactions. This can be understood by the fact that the dynamics is constrained to take place in a small part of the total available Hilbert space. In particular, in the case of infinite-range interactions, the system is described by the  Lipkin-Meshkov-Glick (LMG) Hamiltonian \cite{vidal_entanglement_2004,wilms_finite-temperature_2012}, where the eigenspace of the model is spanned by relatively few Dicke states. We show that, in this case, the bipartite entanglement is bounded by a constant value, which grows logarithmically with the size of the system. For a large system size, this can be thought of as a mean-field limit, where the dynamics is simple to capture with a small number of basis states. 

Finally, we discuss specific experimental parameters for the realization of different regimes in ion traps with finite chain lengths, and experimental measurement protocols for these effects, creating possibilities for the regimes considered here to be observed in the laboratory. We show that the crossover from linear to logarithmic entanglement growth can be observed also for inhomogeneously decaying interactions. Furthermore, we take typical experimental noise sources into account and show that the observable features are robust against these. The result that long-range interactions do not always give rise to strong entanglement in quench dynamics has implications for the realization of large-scale entanglement in quantum simulations in general systems with long-range interactions.

This paper is organized as follows. In
Sec.~\ref{sec:model} we introduce the setup and the model, as well as the entanglement measures we compute. In Sec.~\ref{sec:results} we show how the entanglement growth depends on the model parameters and how
the entanglement distribution mechanisms can be understood. In
Sec.~\ref{sec:experiment}, we show entanglement growth for typical experimental parameters with inhomogeneously decaying interactions and how the entanglement behavior can be measured in noisy experiments. Finally,  in Sec.~\ref{sec:conclusion} we provide a conclusion
and an outlook.

\section{Model for a quench with long-range interactions} \label{sec:model}

In this paper, we study the nonequilibrium dynamics of spatial entanglement in systems with long-range interactions, especially as they are realizable with variable range in ion traps. In this section, we introduce the long-range transverse Ising model governing the time evolution, and the measures of entanglement we compute.

\subsection{Transverse Ising model}

We consider the transverse Ising model with long-range interactions, described by the Hamiltonian
\begin{align} \label{eq:ising_ham} 
\hat H=\sum_{i<j} J_{i,j} \hat \sigma_i^{x} \hat \sigma_j^{x}  + B \sum_i
\hat \sigma_i^{z}.
\end{align} 
Here, the $\hat \sigma^{\alpha}_i$ denote the local
 Pauli matrices ($\alpha=x,z$), $J_{i,j}$ is a general interaction matrix with potentially
long-range interactions, and $B$ is the transverse field. This Hamiltonian can be realized experimentally, e.g., with a string of trapped ions that are harmonically confined in a linear trap, as depicted in Fig.~\ref{fig:fig1}. Using two stable (or metastable) electronic states of these ions as local spin representations at site $i$, $\ket{\!\!\us}_i$ and $\ket{\!\!\ds}_i$, it has been shown \cite{porras_effective_2004} that one can use collective couplings of these local states to motional degrees of freedom of the whole chain to produce the effective spin-model \eqref{eq:ising_ham}  [an example of $J_{i,j}$ for the ion trap experiment``case {\bf B}'' of Sec.~\ref{sec:experiment}, is shown in Figs.~\ref{fig:fig1}(b) and \ref{fig:fig1}(c)]. Note that, throughout this paper we will deal with open boundary conditions, which are typical in ion-chain experiments.

We define the local eigenstates of  $\hat \sigma^{z}_i$ as $\ket{0}_i\equiv \ket{\!\!\ds}_i$ and $\ket{1}_i\equiv \ket{\!\!\us}_i$ with eigenvalues $-1$ and $1$, respectively. We consider a quench experiment [see Fig.~1(a)], where the system starts in the
fully polarized state $|\psi_0 \rangle = \prod_i^M \ket{0}_i$, which is the ground state for $B(t=0)\to \infty$. We are interested in the nonequilibrium dynamics of the many-body quantum state under a coherent evolution, i.e., ($\hbar=1$):
\begin{align}
\ket{\psi_t}={\rm e}^{-{\rm i} \hat H t} |\psi _0 \rangle 
= \sum_{i_1, i_2, \dots i_M} c_{i_1,i_2,\dots, i_M}(t) \prod_k \ket{i_k}_k,
\end{align}
with $i_k\in\{0,1\}$.

We will concentrate  on the case $J_{i,j}>0$ for all $i,j$. However, we note that, nevertheless, we obtain solutions for both the ferromagnetic and the antiferromagnetic case.  Since we start in a state with a real probability amplitude in the spin basis, the evolution for any observable $\hat O_z$ (with $\hat O_z^\dag=\hat O_z$), such as a density matrix of any subsystem, is completely symmetric under the time-reversal transformation $t \leftrightarrow -t$. This can be seen by the fact that
\begin{align}
\bra{\psi_t} \hat O_z \ket{\psi_t}
&= \bra{\psi_0} \cos\left(\hat H t\right) \hat O_z \cos\left(\hat H t\right) \ket{\psi_0}
\nonumber \\
&+\bra{\psi_0} \sin\left(\hat H t\right) \hat O_z \sin\left(\hat H t\right) \ket{\psi_0} 
\nonumber\\
&=\bra{\psi_{-t}} \hat O_z \ket{\psi_{-t}},
\end{align}
where the cross terms have to vanish because of the real coefficients of the initial state and the fact that the expectation value must be real.
 Thus, the evolution of any observable is identical under both Hamiltonians $\hat H$ and $-\hat H$. Therefore, the results we obtain for $B>0$ with the antiferromagnetic ($J_{i,j}>0$) model are identical to the ferromagnetic model ($J_{i,j}<0$) with either a negative field $-B$ or a rotated initial state.
 
In this paper, we will show how the entanglement growth behavior changes with the strength of the magnetic field and the range of the interactions. Initially, we will idealize the interaction matrix, taking the form
\begin{align} \label{eq:interaction_matrix}
J_{i,j} = \frac{\bar J}{|i-j|^\alpha},
\end{align} 
where $\bar J$ denotes the nearest-neighbor interaction strength and $\alpha$ is the decay exponent. This gives a good representation of the basic behavior of the interactions, but in real experiments, there are typically small deviations from the purely algebraic behavior of the interactions. In Sec.~\ref{sec:experiment} we will consider a full interaction matrix $J_{i,j}$ for real experimental parameters, as well as the effects of noise in the experiment.

\subsection{Spatial entanglement}
 
In characterizing the growth of spatial entanglement in the spin chain, we will make use of two complementary measures: The von Neumann entropy for a bipartite splitting of the chain in the center of the system, and the quantum mutual information between two distant spins. The former gives a measure of the overall entanglement buildup, and it also gives an idea of the complexity of the state being generated. The latter measure will give more detailed information as to how correlations propagate spatially, and it will also help us to characterize what part of the entanglement buildup is due to propagation of quasiparticles produced in the quench and which part is due to direct interactions through long-range interactions. Both of these measures are accessible in experiments, though the mutual information is substantially less costly to measure (see Sec.~\ref{sec:measurement} for more information).

\subsubsection{Half-chain von Neumann entropy}
 
Consider a chain of $M$ spins as depicted in
 Figs.~1(a). We can split this system into two halves, $L$ and $R$, in the center of
 the system. In the case that the (pure) state of the composite system
 $\ket{\psi}$ cannot be written as a product state of two states on the
 subsystems $L$ and $R$, i.e., $\ket{\psi}\neq \ket{\psi_A} \ket{ \psi_B }$,
 we call the state entangled. The reduced density
 matrix of the subsystem $L$ is defined via $\rho_L\equiv {\rm tr}_R (
 \ket{\psi} \bra{\psi} )$, where ${\rm tr}_R$ denotes the partial trace
 over the system $R$. This density matrix will only be pure for a product
 state, and in the case of an entangled state, the amount of bipartite
 entanglement is quantified by the von Neumann half-chain entropy of this matrix,
 which is defined as 
 \begin{align} S_{\rm vN}\equiv S(\rho_L) \equiv -{\rm tr}
 \left( \rho_L \log_2 \rho_L \right). 
 \end{align}

The time-dependent growth of the half-chain entropy summarizes the buildup of quantum correlations between two halves of the system. In a sense, it also underlies the complexity of numerical simulations of the quench when using matrix product state (MPS) representations.  As we show in Appendix \ref{app:numerics}, the size of the MPS, represented by the bond dimension $D$, has to grow exponentially as a function of time in the case where $S_{\rm vN}$ grows linearly. Regimes of linear entanglement growth in time are thus important in demonstrating the power of a quantum simulator since realizing such regimes is a necessary requirement in order to observe dynamics that cannot be captured by state-of-the-art numerical techniques over long time scales \cite{trotzky_probing_2012,schuch_entropy_2008}.

\subsubsection{Quantum mutual information}

An alternative measure, which gives more information on the distance of correlations, is the quantum mutual information between two distant spins $i$ and $j$. In an experiment, this is also more straightforward to measure than the von Neumann entropy for a bipartite splitting into two large blocks (see Sec.~\ref{sec:measurement}), and it clearly allows one to distinguish different regimes of entanglement growth. The quantum mutual information is defined as  
\begin{align}
I_{i,j}=S_{\rm vN}(\rho_i) + S_{\rm vN}(\rho_j) - S_{\rm vN}(\rho_{ij}).
\end{align}
Here, $\rho_i={\rm tr}_{k\neq i} (|\psi\rangle\langle\psi|)$ and $\rho_j={\rm tr}_{k\neq j} (|\psi \rangle\langle\psi|)$ denote the reduced density matrices of the single spins (obtained by tracing over all other spins $k$), and $\rho_{ij}={\rm tr}_{k\neq i,j} (|\psi\rangle\langle\psi|)$ is the reduced density matrix of the composite system of the two spins.

Note that one has to be careful when interpreting the half-chain entropy and the quantum mutual information in an experiment in which the quantum state of the whole chain is, in general, mixed because of  coupling to the environment and classical noise. In general, the von Neumann entropy for each reduced density matrix is expected to increase compared to the zero-temperature case \cite{pichler_thermal_2013}. We will consider these imperfections in Sec.~\ref{sec:experiment}.

\section{Entanglement growth dynamics}
\label{sec:results}

In this section, we study the evolution of the entanglement after the quench. We identify three very different regimes: (i) For relatively short-range interactions $\alpha \gtrsim 1$ (depending on the system size), we find a linear growth of the half-chain entropy as a function of time,  which we can understand in terms of free quasiparticle propagation within an effective Lieb-Robinson light-cone. (ii) For long-range interactions $\alpha \sim 0.8,0.9,1$, we find a regime where the half-chain entropy grows logarithmically.  (iii) For nearly infinite-range interactions with $\alpha \lesssim 0.2$, we find rapid oscillations of the half-chain entropy around small values, which we can understand in an effective Dicke-state model \cite{dicke_coherence_1954}. We treat case (i) in Sec.~\ref{sec:short_ranged}, case (ii) in Sec.~\ref{sec:long_ranged}, and case (iii) in Sec.~\ref{sec:longest_ranged}.  In Sec.~\ref{sec:gs_phase_diag}, we show how the entanglement growth in regime (i) depends on the transversal field $B$ and we discover a connection between the entanglement growth rate and the underlying ground-state phase diagram of the model.

  \begin{figure*}[tb] 
\begin{center}
\includegraphics[width=0.95\textwidth]{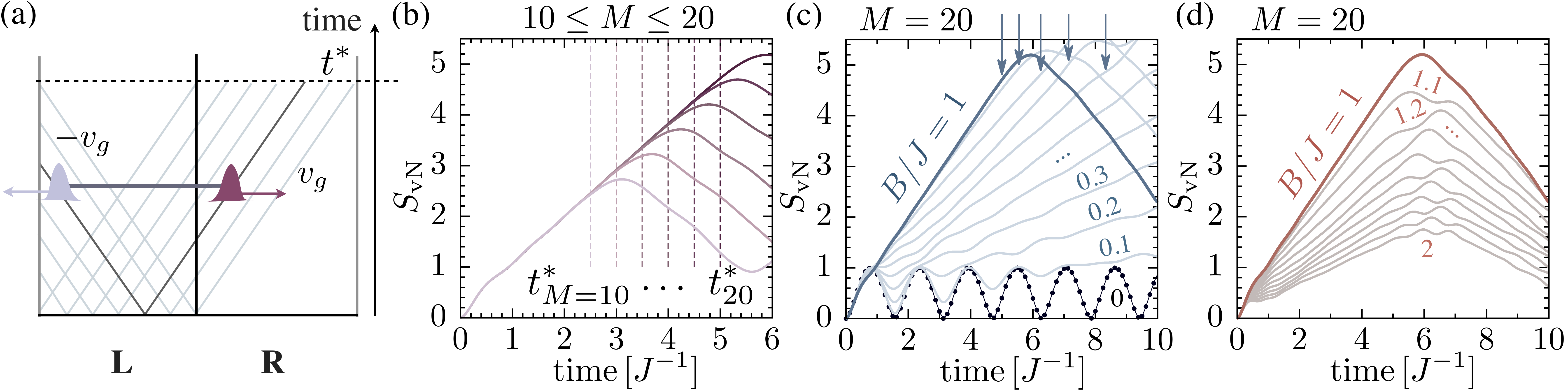} 
\end{center}
\caption{Entanglement growth after a quantum quench in the transverse Ising model in which nearest-neighbor interactions are introduced suddenly. (a) Illustration
  of entanglement distribution, via entangled quasiparticle pair
  excitations which move within a Lieb-Robinson light-cone. Boundary effects for this system with open boundary conditions stop the linear increase at a critical
  time $t^*$. (b--d) Time evolution of half-chain entropies  for $M=10,12,14,16,18$, and $20$
  spins (ED calculation). (b) Boundary effects  as a breakdown of the linear 
   growth. Respective critical times calculated for the free
  fermion model are shown as vertical lines. (c) The crossover from the oscillatory
  behavior for $B=0$ (dots: analytical result) to a linear increase (M=20). With decreasing $B$, boundary effects shift to later times; critical times are indicated as vertical arrows. (d) The half-chain entropy
  growth, which is fastest for $B=1$ and decreases again for $B>1$ (M=20). \label{fig:nn_results}}
\end{figure*}

\subsection{Entanglement dynamics for relatively short-range interactions ($\alpha \geq 1$)}
\label{sec:short_ranged}

\subsubsection{Nearest-neighbor interactions} 

To understand the entanglement entropy growth behavior in this regime,
it is instructive to first revise the case of nearest-neighbor
interactions, i.e., an interaction matrix \eqref{eq:interaction_matrix}
with a decay exponent $\alpha\rightarrow\infty$, and discuss the dynamics
of the quantities we study here. In this limit, the model
Hamiltonian \eqref{eq:ising_ham} becomes a standard transverse Ising
model of the form
\begin{align}
    \label{eq:ising_nn_ham}
    H=J \sum_{i} \sigma^x_i \sigma^x_{i+1}+B\sum_i \sigma_i^z
    ,
\end{align}
which has been well studied in the literature. Note that since the spectrum of the Hamiltonian is symmetric under the exchange $B \leftrightarrow -B$, the dynamics will not only be identical under a change of the sign of the total Hamiltonian, but also under a change of the sign of $B$, and we therefore focus on $B>0$ here. The model \eqref{eq:ising_nn_ham} can be diagonalized analytically
\cite{pfeuty_one-dimensional_1970} (see Appendix \ref{app:analytics} for
more details). After performing a Jordan-Wigner transformation and
diagonalizing the quadratic Hamiltonian in quasimomentum space, the
resulting diagonal model is a model of free fermions $\gamma_q$,
\begin{align}
\label{eq:ising_nn_ham_diagonal}
H=\sum_{q} {\epsilon}_q \left(\gamma_q^\dag \gamma_q - \frac{1}{2} \right),
\end{align}
The $\gamma_q$ ($\gamma_q^\dag$) are the annihilation (creation)  operators for a fermionic quasiparticle with quasimomentum $q$, which obey the anticommutation relations $\{\gamma_q, \gamma_p^\dag
\}=\delta_{q,p}$. In the thermodynamic limit, i.e., for a chain of infinite length, the quasimomenta become
continuous $-\pi < qa < \pi$ ($a$ is the spatial separation between
the spins), and the dispersion relation of the free particles is
two fold degenerate for $q=-q\neq 0$ and given by $\epsilon_q= 2
\sqrt{(J-B)^2+4JB \sin^2(qa/2)}$. The group velocity of quasiparticle excitations in this system is given by
$v_g(q)/a= d\epsilon_q/d(qa)$. The maximum velocity of the quasiparticles gives rise to the Lieb-Robinson bound, which defines an effective light cone for spatial correlations, outside of which the correlations are exponentially suppressed \cite{lieb_finite_1972}. This sets an upper linear bound on the block entropy growth, as we will see below. It is straightforward to calculate that
the fastest particles move at a Lieb-Robinson velocity $v_R= \max |v_g| = 2 a J$ for $B \geq
J$ and $v_R = 2 a B$ for $B<J$.

Following \cite{calabrese_evolution_2005}, we can understand the
entanglement distribution mechanism in model \eqref{eq:ising_nn_ham_diagonal}  as follows: In
a coherent time evolution, the initially excited state acts as a
source for quasiparticle excitations. Pairs of the free fermions with
quasimomenta $p$ and $-p$, which have been created at a certain point
in space, are entangled pairs. These pairs move freely through the
system with  corresponding group velocities $v_g$ and $-v_g$,
respectively. Parts of pairs that have been produced in block $L$ and
arrive in block $R$ entangle the two blocks. An illustration of this
mechanism is given in Fig.~\ref{fig:nn_results}(a). Thus, the arrival rate in block $R$ for quasiparticles belonging to a pair created in
block $L$ is constant. Therefore, the increase of
half-chain entropy is linear, and we expect $S_{\rm vN}=\eta v_g t$, with some constant $\eta$. Since the group velocity is limited by the Lieb-Robinson bound,  $S_{\rm vN}\leq \eta v_R t$.

We can test this mechanism explicitly by making use of the boundary
effects with open boundary conditions. Consider a quasiparticle pair, which has been created at the
left edge of block $L$ and moves at the Lieb-Robinson
velocity. As soon as the right-moving quasi-particle arrives in block
$R$, the linear entanglement increase has to break down since there are
no more entangled pairs available to the left that could further
entangle blocks $L$ and $R$. We can estimate the time at which this happens
as $t^*=(M/2)/v_R$, which corresponds to $t^*=(M/4J)$ for $
B\geq J$ and to $t^*=(M/4B)$ for $ B < J$.  In
Fig.~\ref{fig:nn_results}(b), we plot a comparison of this critical time
with a numerical exact diagonalization simulation (ED; see Appendix \ref{app:numerics} for details) of the half-chain entropy evolution for
$B=J$ for increasingly large system sizes of $10 \leq M \leq 20$
spins. We see that, as expected exactly at the critical time, for each
system size, the entropy starts to level off and remarkably
reduces again after this maximum peak.

The quench experiment for $B=J$ is special in the sense that this is a quench to 
the critical point of the quantum phase transition of Hamiltonian
\eqref{eq:ising_nn_ham}. We will now ask how the entanglement growth
depends on $B$. In the limit of $B\to
\infty$, the initial state becomes an eigenstate of the system and no
evolution will take place, i.e., $S_{\rm vN}(t)=0$. On the other hand, in the limit of $B \to 0$, the Hamiltonian
has a spectrum with $M$ degenerate levels, which are separated by an
energy of $\sim 2J$ (spin flips). Thus, in the latter case, we expect
dynamics which is dominated by oscillations between those levels at a
frequency scale given by $J$. Indeed, for $B=0$ it is straightforward
to calculate analytically that $S_{\rm
  vN}(t)= -\cos^2(Jt)\log_2[\cos^2(Jt)]
-\sin^2(Jt)\log_2[\sin^2(Jt)]$. In Fig.~\ref{fig:nn_results}(c), we find
the expected behavior for $B<J$ in an exact diagonalization simulation of a
system with $20$ spins. For increasing $B$, the oscillatory behavior of $S_{\rm vN}$
(which fits the analytical result) breaks down and changes into a
linear increase before boundary effects become important. Since the
Lieb-Robinson velocity decreases with $B$ for
$B<J$, correspondingly the boundary effects shift to later times for
smaller $B$ [critical times $t^*$ are indicated as vertical arrows in Fig.~\ref{fig:nn_results}(c)]. It is interesting to note that in this case the maximum value of $S_{\rm vN}$ can actually be larger than for $B=J$. In Fig.~\ref{fig:nn_results}(d), we analyze the opposite case of
$B>J$. Remarkably, we find that, also away from the critical point the half-chain entropy growth becomes slower with increasing $B$. Below, we will find that this also holds for finite-range interactions with $\alpha \gtrsim 2$ and that the fastest entanglement growth precisely follows  the point of the phase transition.

\begin{figure*}[t] 
\begin{center}
\includegraphics[width=0.95\textwidth]{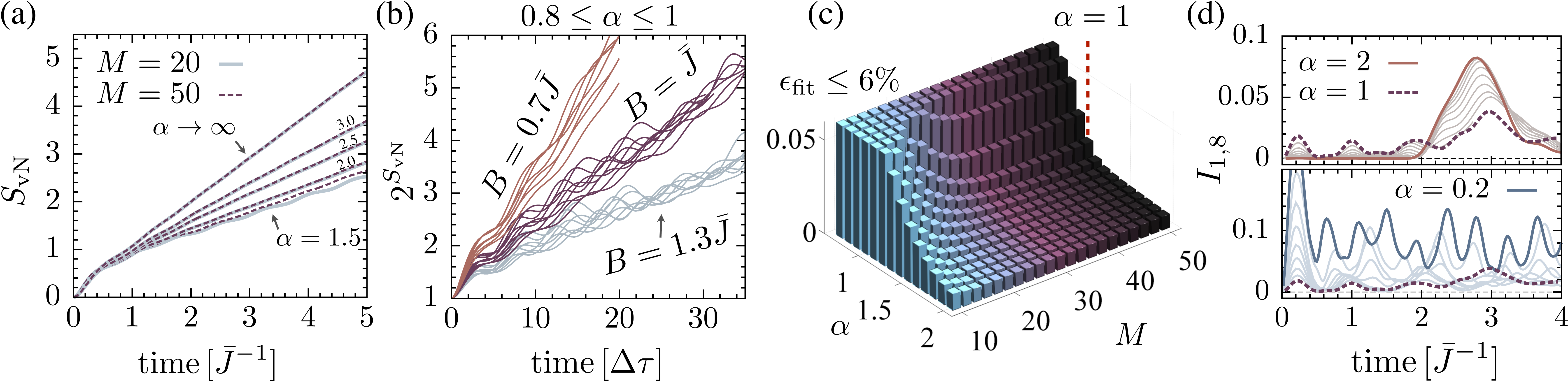} 
\end{center}
\caption{Entanglement growth after a quantum quench in the transverse Ising model in which algebraically decaying interactions are introduced suddenly. (a) Time-evolution of the half-chain entropy after the quench for $B=1$ and varying decay exponents $\alpha=1.5,2,2.5,3$, and $\infty$ (from bottom to top).  Solid lines are ED results for $M=20$ spins; dashed lines are MPS--MPO results for $50$ spins (converged with MPS bond dimension $D=192$). For $\alpha \geq 2$, the growth is clearly  linear and independent of the system size. (b) Time evolution of $2^{S_{\rm vN}}$. Each of the three bundles of lines contains the results for $M=30,40,50$ spins and $\alpha=0.8,0.9,1$ (MPS--MPO simulation, converged with $D=192$). On top of the oscillations, the growth is logarithmic (straight line on the exponential scale). Time is given in units of the inverse Hamiltonian norm, $\Delta \tau$ (cf.~Sec.~\ref{sec:long_ranged}).  (c) Finite-size scaling of the crossover from linear  $S_{\rm vN}$ growth to a logarithmic one visualized by the error of a linear fit, $\epsilon_{\rm fit}$,  in the interval $1 < t\bar J < 3$ as a function of $\alpha$ and $M$ ($B=1$, ED \& MPS--MPO simulations, $D=192$). For large systems, the crossover occurs around $\alpha\sim1$. (d) Time evolution of the mutual information between spins $1$ and $8$, $I_{1,8}$ ($M=20$, $B=\bar J$, ED). The upper panel shows results for $2 \geq \alpha \geq 1$, the lower panel for $1 \geq \alpha \geq 0.2$. The signature of linear growth of the half-chain entropy is the arrival of a quasiparticle peak after a certain time, whereas for $\alpha\lesssim 1$, distant spins become entangled instantaneously.  \label{fig:crossover}}
\end{figure*}

\subsubsection{Finite-range interactions ($\alpha\gtrsim1$)} 

We now investigate the situation of relatively short-ranged interactions, which extend beyond nearest neighbors and decay algebraically with $\alpha \gtrsim 1$. In Fig.~\ref{fig:crossover}, we demonstrate that the picture of entanglement distribution via entangled quasiparticle pair propagation also holds for a large range of finite $\alpha \gtrsim 1$ (depending on the system size). We find that, despite the existence of direct spin-spin interactions over all distances, it is the propagation of quasiparticles that dominates the dynamics of entanglement growth over this range of $\alpha$ values. 

One marked signature of the linear half-chain entropy growth is that, before boundary effects become important, the rate of the growth is essentially independent of the size of the system, as shown in Fig.~\ref{fig:crossover}(a). In this figure, we show the time evolution of $S_{\rm vN}$ after the quench for various decay exponents in the range $\infty \geq \alpha \geq 1.5$. The solid lines show an ED simulation for a system of $20$ spins, and the dashed lines are for $M=50$ spins. We obtain the results for large systems using t-DMRG methods. Specifically, we use a matrix product operator (MPO) \cite{crosswhite_applying_2008,pirvu_matrix_2010,frowis_tensor_2010} of the Hamiltonian to time evolve a  MPS via a Runge-Kutta-type method \cite{garcia-ripoll_time_2006} (see Appendix \ref{app:numerics} for more details). In all cases with $\alpha > 2$, we find that the two lines coincide and that the increase is linear. For $\alpha=1.5$, we find a slight change in the behavior in the sense that the results for $M=20$ and $M=50$ start to differ. When we further increase the range of interactions, we find that this linear growth changes to a logarithmic one, as we demonstrate in Fig.~\ref{fig:crossover}(b) using a t-DMRG calculation with $\alpha=0.8,0.9,1$ and for $M=30,40,50$. We treat this case in more detail in Sec.~\ref{sec:long_ranged}.

In Fig.~\ref{fig:crossover}(c), we show an overview over the regime of linear half-chain entropy growth and its finite-size scaling. Therefore, as a function of system size and $\alpha$, we plot the error of a linear fit,  $\epsilon_{\rm fit}$, which is defined as a $95\%$ confidence interval on the slope coefficient and it is cut off at $6\%$.  In large systems we find that the linear increase (small error) breaks down at $\alpha\sim1$. For smaller systems, boundary effects and finite-size effects become significant, and the linear regime breaks down at larger $\alpha$. Note that the change in behavior for large systems at $\alpha \sim 1$ can, in a sense, be understood since this marks the point at which, in the thermodynamic limit, the sum in the interaction term in the Hamiltonian begins to diverge with increasing system size.

We can also identify the regime of linear  growth of $S_{\rm vN}$ by looking at the mutual information between distant spins. In the upper panel of Fig.~\ref{fig:crossover}(d), we plot the time evolution of the mutual information $I_{1,8}$, between sites $1$ and $8$, for $1\leq \alpha \leq 2$ in a system of $20$ spins and for $B=\bar J$. As a clear signature of the regime of linear growth of the half-chain entropy, we find that the mutual information remains nearly zero for a certain time until it suddenly peaks at a time corresponding to the arrival of an "entangling" quasiparticle pair originally produced on a site between the two spins. For nearest-neighbor interactions, this arrival time is consistent with the analytically calculated Lieb-Robinson velocity (cf. Appendix \ref{app:analytics}), and we find that the same mechanism still holds for rather long-ranged interactions of $\alpha\sim 2$. In contrast, for the regime of logarithmic growth of $S_{\rm vN}$, we find a markedly different behavior [lower panel of Fig.~\ref{fig:crossover}(d)], which is discussed in the next section.

We emphasize that the fact that the entanglement growth mechanism is directly reflected in the time dependence of  the mutual information between two distant spins is very important for experimental observations. Instead of having to reconstruct $2^{(M/2)}\times 2^{(M/2)}$ density-matrix elements of a large block via quantum-state tomography, the growth behavior of the half-chain entropy can be directly verified by measuring only $4\times4$ density matrices for a system of two composite spins.  In Sec.~\ref{sec:measurement}, we will show how the measurement further simplifies for the particular quench we consider here.

\subsection{Entanglement dynamics for long-range interactions}

In this section, we study the entanglement growth for very long-range interactions with $\alpha \leq 1 $. In this regime, the picture of entangling quasiparticles that move freely within a light cone breaks down, and instead distant parts of the system can become almost instantaneously entangled based on direct interactions. We observe that for $\alpha \sim 0.8,0.9,1$, the half-chain entropy can still increase steadily as a function of time for our quench, but that the increase becomes logarithmic instead of linear. When further increasing the range of interactions for $\alpha \lesssim 0.2$, we find a regime where $S_{\rm vN}$  oscillates rapidly around small values. We understand this behavior via an effective model in a basis of Dicke states  \cite{dicke_coherence_1954} for infinite-range interactions $\alpha=0$.

\subsubsection{Logarithmic entropy growth}
\label{sec:long_ranged}

When increasing the range of interactions, eventually the linear growth of $S_{\rm vN}$  breaks down, and the growth becomes logarithmic, as shown in Fig.~\ref{fig:crossover}(b). For very long-range interactions, the time scale of the dynamics is dominated by the interaction-energy term in the Hamiltonian. Thus, to make a valid comparison, it is favorable to measure the time in inverse units of the matrix norm instead of $\bar J$. For Hamiltonian \eqref{eq:ising_ham}, we can calculate the Frobenius norm as $\| H \|= 2^{M/2} \sqrt{\left(\sum_{j>i}^M J_{i,j}^2 + M B^2 \right)}$. The $2^{M/2}$ prefactor is due to the exponential growth of the Hilbert space with $M$, and we define the time unit per spin realistically as $\Delta \tau^{-1}=\| H \| 2^{-M/2}$.  In Fig.~\ref{fig:crossover}(b), we plot the evolution of the von Neumann entropy in these units for system sizes of $M=30,40,50$, for $\alpha=0.8,0.9,1$ and for $B=0.7 \bar J,1  \bar J,1.3  \bar J$. If, ideally, the entropy increases logarithmically without oscillation, $S_{\rm vN}=\log_2(C t+1)$, with some constant $C$. On an exponential scale, i.e., by plotting $2^{S_{\rm vN}(t)}$, we would see a straight line. In Fig.~\ref{fig:crossover}(b), we indeed find oscillations around a straight line. It is remarkable that, for a fixed value of $B$, independently of the system size and for all $\alpha=0.8,0.9,1$, we find roughly the same constant $C$ in units of $\Delta \tau^{-1}$. With decreasing $B$, i.e., for quenches that put an increasing amount of energy into the system, the constant increases. For interactions with $\alpha \lesssim 0.7$, the oscillations become more dominant so that the logarithmic increase is hard to verify. 

Also, the time evolution of the mutual information between distant spins shows a completely different qualitative behavior for $\alpha \lesssim 1$ than for $\alpha \gtrsim 1$. In Fig.~\ref{fig:crossover}(d) we show the evolution of $I_{1,8}$ after the quench for a system of $20$ spins ($B=\bar J$) for $2 \geq \alpha \geq 0.2$. In the upper panel of Fig.~\ref{fig:crossover}(d), we find that the incoming-wave picture breaks down when $\alpha$ decreases from $\alpha=2$ to $\alpha=1$. In the latter case, we find a mixed behavior, where a wave peak is still roughly visible around $t \bar J \sim 3$; however peaks also appear for very short times. These peaks indicate that, because of the long-range part of the interaction, distant parts of the system become rapidly entangled with an immediate increase in correlations after the quench. When further increasing the interaction range [lower panel of Fig.~\ref{fig:crossover}(d)], these contributions to $I_{1,8}$ become dominant and the quasiparticle peak disappears completely. While for $\alpha \sim 1$, $I_{1,8}$ still shows some slow overall increase as a function of time, in the case of nearly infinite-range interactions ($\alpha \lesssim 0.2$), we only find rapid oscillations around a constant value of $I_{1,8}\sim 0.1$. In an experiment, the decrease in the height of the short-time peak of the mutual information could be used as an indicator for the crossover from the logarithmic half-chain entropy growth regime to the linear one. Furthermore, we find that this height, in contrast to the frequency of the oscillations, is independent of the system size.

\subsubsection{Entanglement dynamics for infinite-range interactions}
\label{sec:longest_ranged}

To understand the rapid oscillations and small half-chain entropy for decay exponents of $\alpha \lesssim 0.2 $, it is instructive to consider the
case of infinite-range interactions, i.e., $\alpha=0$. In this limit, each spin interacts
with equal strength with all others and the (``mean-field'') Hamiltonian reads
\begin{align} 
\label{eq:infrange_ham} 
\hat
H= J \sum_{i<j}^M \hat \sigma_i^{x} \hat \sigma_j^{x}  + B \sum_i^M
\hat \sigma_i^{z}.
\end{align} 
As in the nearest-neighbor case, this limit is analytically exactly solvable.   We can
introduce effective spin-$M/2$ operators $S_{x,y,z}\equiv \sum_i^M
\hat \sigma_i^{x,y,z}$. With these, we can rewrite Hamiltonian
\eqref{eq:infrange_ham} as
\begin{align} 
\label{eq:dicke_ham} 
\hat
H= \frac{J}{2} S_x^2 + B S_z - \frac{J}{2} M.
\end{align} 
In the literature, this model is well known as the Lipkin-Meshkov-Glick (LMG) model, and its entanglement properties have been studied, e.g., in \cite{vidal_entanglement_2004,wilms_finite-temperature_2012}. A basis of Hamiltonian \eqref{eq:dicke_ham} is given by Dicke spin-M/2 states, which are defined as
\begin{align}
\left|S=\frac{M}{2},m_S=n_1-\frac{M}{2}\right\rangle \equiv \mathcal{S} |\{n_0,n_1 \} \rangle.
\end{align}
Here, $n_0, n_1$ denote the number of spins (down and up,
respectively), and $\mathcal{S}$ is the symmetrization operator. For example, for a system of four spins, we would have an effective spin-2 model, and one particular Dicke state would be $\ket{S=2,m_S=-1}=(\ket{0001} + \ket{0010} +
\ket{0100} + \ket{1000})/2$. In this picture
the quench experiment is equivalent to a free evolution under
Hamiltonian \eqref{eq:dicke_ham} with the initial state
$\ket{\psi_0}=\ket{S=M/2,m_S=-M/2}$.

\begin{figure}[b] 
\begin{center}
\includegraphics[width=0.45\textwidth]{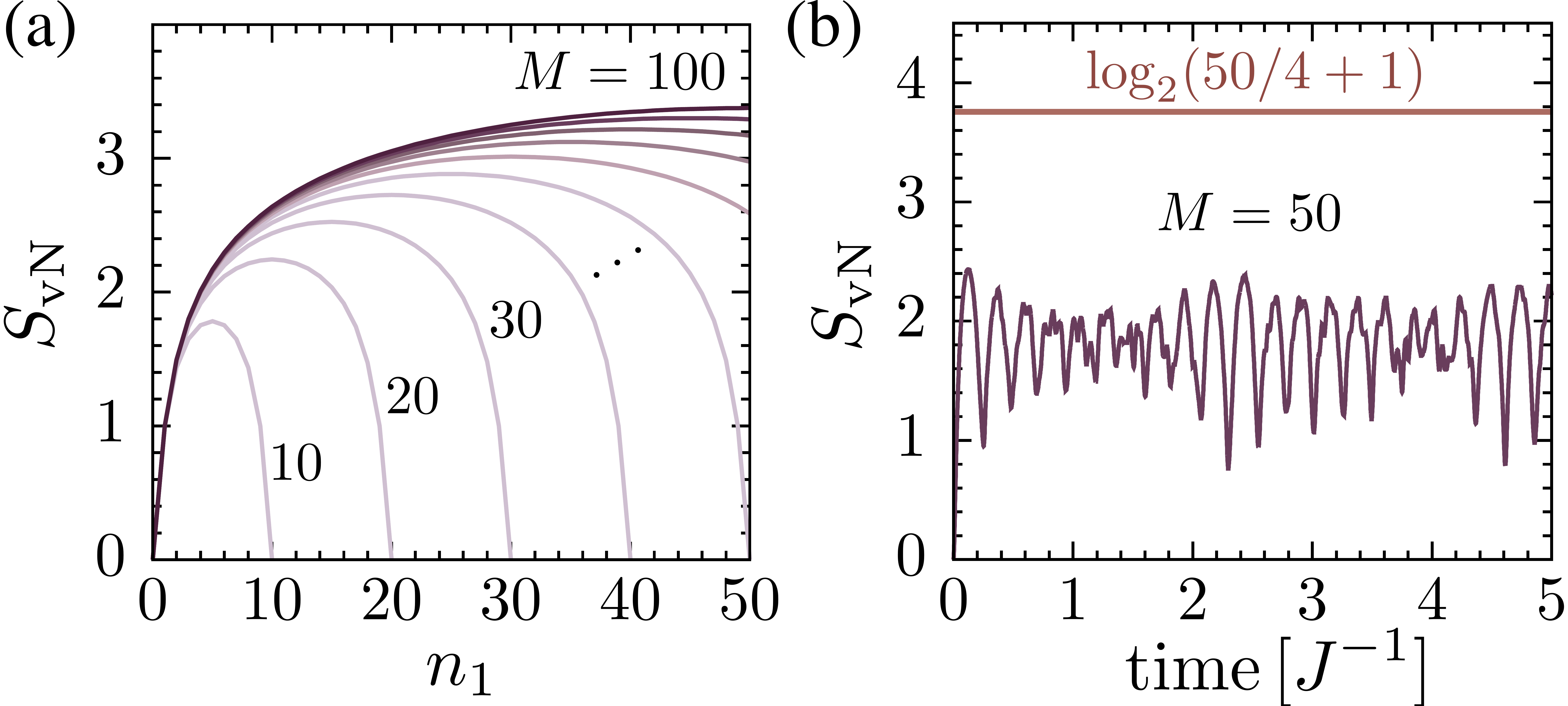} 
\end{center}
\caption{ (a) The half-chain entropy for single Dicke states
  as a function of the number of spins up and for system sizes $M=10, \dots, 100$. (b) The time evolution of the
 half-chain entropy in our specific quench experiment for a chain of $50$ spins. As shown in the text, the entropy is bounded by a constant
  $S_{\rm vN}\leq\log_2(M/4+1)$. \label{fig:dicke_entanglement}}
\end{figure}

It is straightforward to calculate the half-chain von Neumann
entropy for an arbitrary Dicke
state \cite{latorre_entanglement_2005} (see Appendix \ref{app:analytics} for more details). One finds that 
\begin{align}
\label{eq:dicke_vne}
S_{\rm vN}&=-\sum_l p_l \log_2(p_l) \\
p_l(n_1) &= \frac{ {M/2\choose l} {M/2\choose n_1-l} }{ {M \choose n_1}},
\end{align}
where $p_l$ are combinatorial factors depending only on the number of
single up spins of the corresponding Dicke state and $0 \leq l \leq
M/2$. Example results are plotted in
Fig.~\ref{fig:dicke_entanglement}(a), and it is important to note that,
simply because the sum in \eqref{eq:dicke_vne} contains a
maximum of $M/2+1$ terms, the entropy is bounded by
$S_{\rm vN}\leq\log_2(M/2+1)$.

In our quench, the time-dependent state will assume a
superposition of Dicke states, $|\psi(t)\rangle=\sum_m c_m(t)
|{M/2,m}\rangle $ with $c_m(t=0)=\delta_{m,-M/2}$. In the small Dicke
Hilbert space, the time evolution can be easily numerically simulated,
and we can, at any instance in time, construct the reduced density
matrix for the left system. From this density matrix, the von Neumann entropy can be readily
calculated, and examples are shown in
Fig.~\ref{fig:dicke_entanglement}(b). Again, we find, simply by noting
that the dimension of the $M/4$ Dicke subspace is $M/2+1$, that for
an arbitrary Dicke-state superposition, the von Neumann entropy
is limited by $S_{\rm vN}\leq \log_2(M/2+1)$. However, for our
specific experiment, a tighter bound can be found. Since the
Hamiltonian \eqref{eq:dicke_ham} only couples spin states with
$m_S=\pm 2$, the time-dependent coefficients $c_m(t)$ can only be
nonzero for states with $m=-M/2,-M/2+2,\dots,M/2$. Assuming an even
number of spins, the entropy is then limited by $S_{\rm vN}\leq
\log_2(M/4+1)$, which is in agreement with the exact von Neumann entropy evolution, as shown in Fig.~\ref{fig:dicke_entanglement}(b).

\subsection{Ground-state phase-transition point and entanglement growth} 
\label{sec:gs_phase_diag}

\begin{figure}[t] 
\begin{center}
\includegraphics[width=0.42\textwidth]{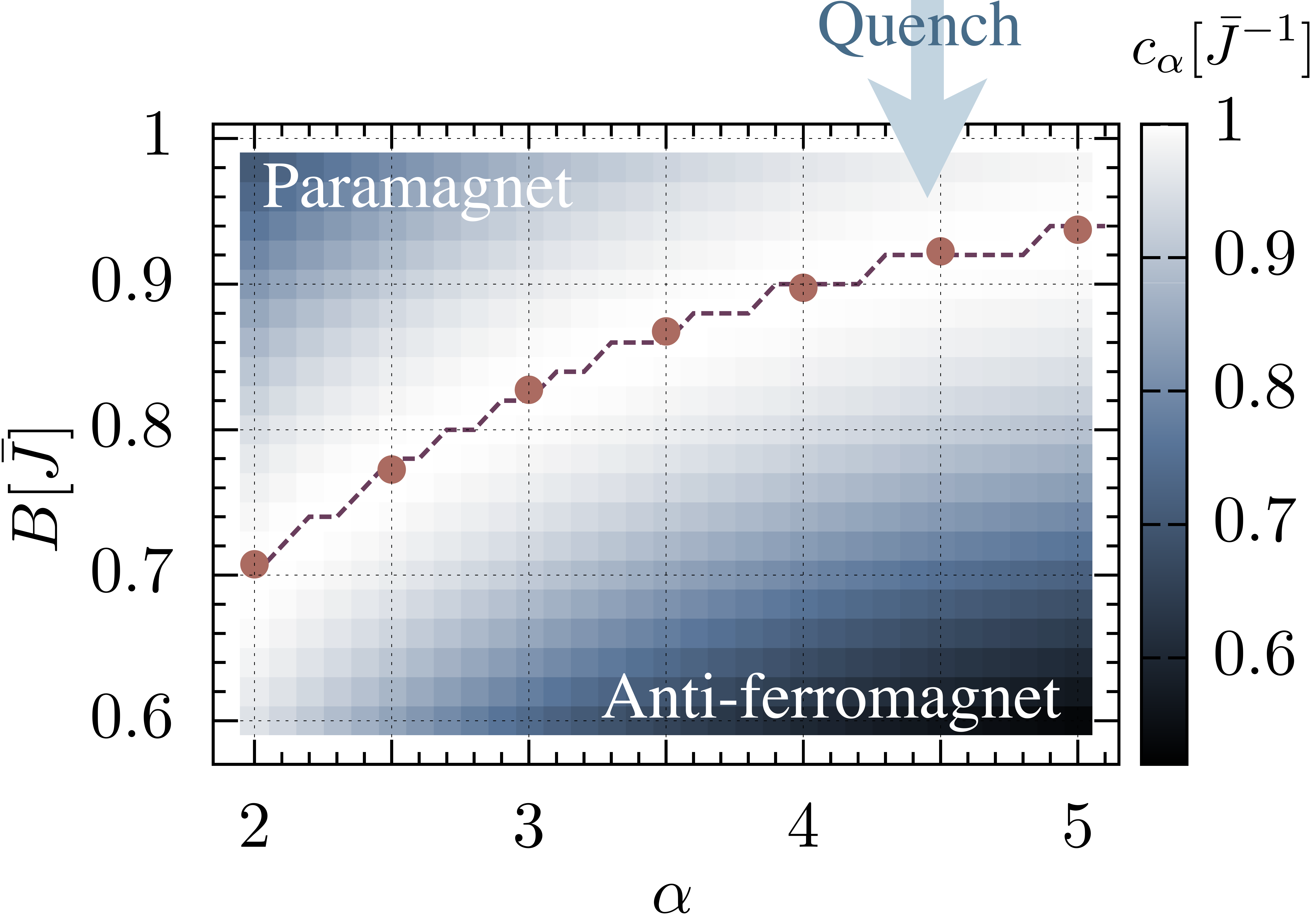} 
\end{center}
\caption{The half-chain entropy growth rate $c_\alpha$ from a linear fit $0 < t \bar J < 3$ after the quench ($M=20$, ED). Each $c_\alpha$ is normalized to its maximum value in the range $0.6<B<1$, $c_\alpha=[dS_{\rm vN}/dt]/{\max[dS_{\rm vN}/dt(B)]}$. The dashed line is the contour for maximum growth, i.e.,~$c_\alpha=1/\bar J$. The points show the location of the quantum phase transition, which we extract from a MPO Lanczos diagonalization for $100$ spins.\label{fig:phase_diagram}}	
\end{figure}

In this section, we study how the linear growth of $S_{\rm vN}$ for $\alpha \leq 2$ depends on the value of  $B$ and the decay exponent $\alpha$. We focus on the case $\bar J=1$, $B>0$. For ground states, as for the nearest-neighbor transverse Ising model, the long-range model undergoes a quantum phase transition from an antiferromagnetic to a paramagnetic phase at a critical field $B_c$ for all decay exponents $\alpha$ \cite{dutta_phase_2001,koffel_entanglement_2012}. For example, for $\alpha=3$ it has been calculated that $B_c\approx0.83$ \cite{deng_effective_2005}. As shown in Ref.~ \cite{deng_effective_2005}, we can estimate the point of the phase transition by locating the discontinuity of the transverse magnetization, i.e., the jump in $d^2 m_z/dB^2$, where $m_z=\sum_m  \langle \sigma^z_m\rangle$. In order to do this for moderately large systems, we use a MPO Lanczos diagonalization for $100$ spins. In Fig.~\ref{fig:phase_diagram}, we compare this ground-state phase-transition point for a moderate system size to the linear growth rates of the von Neumann entropy in a small systems of only $20$ spins as a function of $\alpha$ and $B$. It is remarkable that the point of the ground-state phase transition is not only reflected in the scaling behavior of the block entropy in the ground state \cite{koffel_entanglement_2012}, but we also find that in the evolution following the quench from $B(t=0)=\infty \to B$, the growth rate of the von Neumann entropy as a function of time is largest at the critical points $B_c$. Since the entangling quasiparticles are bounded by the Lieb-Robinson bound, this effect is independent of the system size up to  times $t^*$ when boundary effects limit the quasiparticle propagation.

\section{Entanglement growth and measurement using trapped ions} 
\label{sec:experiment}

\subsection{Entanglement growth for realistic experimental parameters}

\begin{figure}[tb] 
\begin{center}
\includegraphics[width=0.48\textwidth]{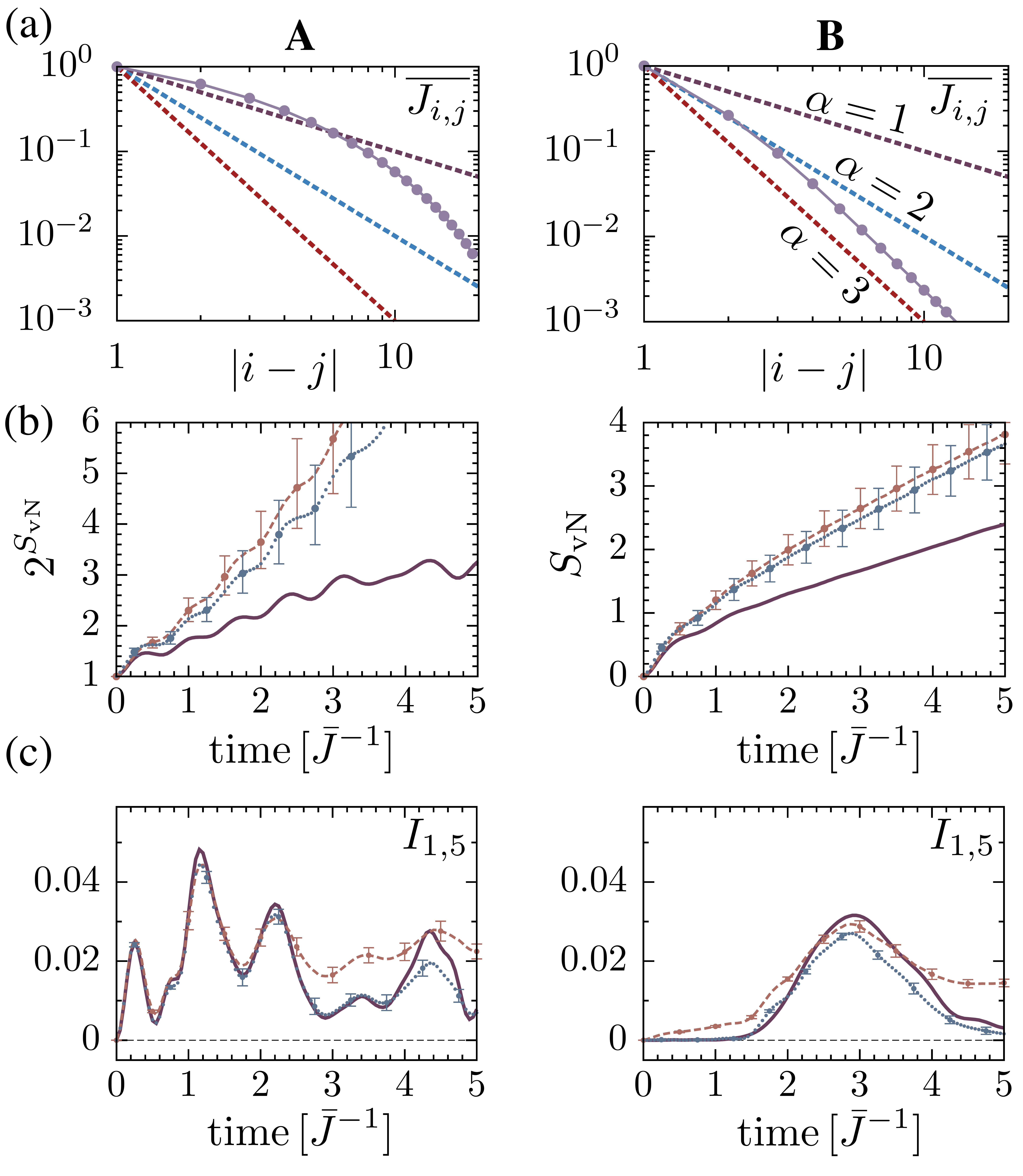} 
\end{center}
\caption{(a) The averaged normalized interactions for the two experimental setups (grey dots). In case {\bf A}, interactions decay with $\alpha <  1$ over short distances; in case {\bf B}, they decay as $\alpha \sim 2$. (b) The evolution of the half-chain von Neumann entropy $S_{\rm vN}$ (exponential scale for case {\bf A}) in a system with $20$ ions (ED). The solid line is the idealized noiseless case. The dashed lines in (b) and (c) are for fluctuations of the magnetic field with $s_1=0.01/ \bar J$, the dotted lines in (b) and (c) are for fluctuations in the coupling strength with $s_2=0.01/\bar J$  ($200$ noise trajectories). (c) Time evolution of the mutual information $I_{1,5}$. The characteristic features of the two entanglement growth regimes survive in the presence of the noise.  \label{fig:real_experiment}}
\end{figure}

In this section, we ask to what extent the effects shown in the previous sections are experimentally observable in ion traps. Therefore, we consider two experimentally realistic full interaction matrices $J_{i,j}$, which show the characteristic behavior of linear entanglement growth and logarithmic growth. In case {\bf A}, over short distances, the averaged interactions decay as $\alpha < 1$ (logarithmic growth regime), and in case  {\bf B}, they decay as $\alpha\sim2$ (linear growth regime), as depicted in Fig.~\ref{fig:real_experiment}(a). We define the energy unit by the largest element of $J_{i,j}$, which we denote $\bar J$ in this section.

We consider a linear chain of 20 ions that interact via the mechanism described in \cite{PhysRevLett.103.120502, 1367-2630-14-9-095024}. In summary, a force is applied that couples the electronic ``spin'' state of the ions to the spectrum of closely spaced (nearly frequency degenerate) vibrational modes transverse to the ion string. By setting the driving force to be far off-resonant the phonon states can be adiabatically eliminated, allowing an analogue simulation. While the simulations presented are for $^{40}$Ca$^+$ ions driven with bichromatic laser fields, similar interaction matrices can be derived in a number of other systems. The exact experimental parameters considered are given in  Appendix \ref{app:exp}.

We show the evolution of the half-chain entropy and the mutual information $I_{1,5}$ in Figs.~\ref{fig:real_experiment}(b) and \ref{fig:real_experiment}(c), respectively. As expected from the  averaged decay of the interactions in both cases {\bf A} and {\bf B}  and from the previous discussions, we find that for the case {\bf B}, $S_{\rm vN}$ increases linearly in time, while for {\bf A}, the growth behavior is logarithmic. Accordingly, we find the behavior of the mutual information as we have found it for the case of homogeneously decaying interactions: In case {\bf B}, the initial mutual information is zero, and a peak appears at a time around $2.8/\bar J$. In case {\bf A}, in the logarithmic regime, we find an instantaneous increase of $I_1$ due to the long-range part of the interactions, which entangles distant parts instantaneously.
                                                                                                    
In a realistic experiment setup, the string of ions will be subjected to noise. Here, we consider the two most significant imperfections: (i) fluctuations in the energy splitting of the electronic states used to encode a spin (e.g., due to ambient magnetic field fluctuations) and (ii) fluctuations in the coupling strength between the spin-dependent force mechanism and the ions (e.g., due to laser-intensity fluctuations). Noise of the form (i) will lead to a correlated rotation of the qubits around the $z$ axis, while noise of type (ii) causes a stochastic fluctuation of the overall interaction strength $\bar {J}$.
We idealize both cases as white noise fluctuations $\xi(t)$, $\overline{\xi(0)\xi(t)}=s_\alpha \delta(t)$ with strength $s_\alpha$ (the bar denotes the time average). 
The evolution of the state can then be described by a Stratonovich stochastic Schr\"odinger equation \cite{gardiner_handbook_2004,pichler_heating_2013}
\begin{align}
{d\ket{\psi}}= -i \hat H \ket{\psi} dt -i \sqrt{s_\alpha}  \hat L_\alpha  \ket{\psi}dW(t)
,
\end{align}
where $dW(t)$ is the Wiener increment, and $\hat L_\alpha$ is the noise (jump) operator. For case (i)  $L_1=B \sum_i \hat \sigma_i^z$; for case (ii) $L_2= \sum_{i,j} J_{i,j} \hat \sigma_i^x \hat \sigma_j^x$. Equivalently, we can derive a master equation for the evolution of the full density matrix,
\begin{align}
\frac{d \rho} {dt}  = -i \left[ \hat H, \rho \right] - \frac{s_\alpha}{2} \left[ \hat L_\alpha, \left[ \hat L_\alpha, \rho \right] \right]
.
\end{align}

In the long-time limit, the master equation drives the system into a state that commutes with the jump operators. For noise of type (i), this is a state diagonal in the $z$ basis, and for type (ii), it is a state diagonal in the $x$ basis. For the time scales of the experiment, the dynamics consists of a complicated interplay between the coherent evolution and the dissipative part. In general, we expect that noise of type (i) leads to a global dephasing in the sense that it will reduce the purity of the full state and thus result in a slightly higher measured entropy, whereas noise of type (ii) can be more complicated because an overall fluctuation of $\bar J$ acts with different strengths between different spins according to $J_{i,j}$. We can simulate the evolution of the master equation numerically by evolving the stochastic Schr\"odinger equation in time using a first-order semi-implicit method with strong order 1.0 convergence \cite{gardiner_handbook_2004} and statistically averaging over a large amount of trajectories.

In Figs.~\ref{fig:real_experiment}(b) and \ref{fig:real_experiment}(c), we find that, as expected, the noise adds an additional entropy growth as a function of time for both the half-chain entropy and the mutual information. However, the underlying entropy features, which arise from the entanglement buildup in the coherent evolution, remain clearly visible. For example, instead of the short-time initial mutual information being zero in the regime of the linear $S_{\rm vN}$ growth regime, for noise of type (i) we find an overall increase of $I_{1,5}$ as a function of time. Nevertheless, the quasiparticle peak clearly remains observable even in the presence of the noise. In general, we find that the overall entropy growth that is induced by the fluctuations on $B$ is larger than the one induced by fluctuations on $J_{i,j}$. In particular, we find that the mutual information between distant spins is very robust against noise on the coupling strength because of the decay of $J_{i,j}$ with distance. In case {\bf A}, we find that for long times, the characteristic logarithmic growth eventually breaks down because of the entropy increase from the noise; however, for $t \bar J \lesssim 3$, it remains observable. In general these results suggest that the mutual information is a very robust experimental measure for the entanglement growth behavior of the systems, which, furthermore, can be easily extracted from experimental data as we show in the next section.

\subsection{Measurements of block entropies and mutual information}
\label{sec:measurement}

We will now briefly review how the entanglement measures we used in this paper can be measured experimentally. To calculate the von Neumann entropy of a subsystem $A$ of $l$ spins (which do not have to be next to each other), one can simply measure the reduced density matrix of that block. While the process of measuring this matrix is known as quantum-state tomography, we show that for our particular experimental setup this tomography simplifies significantly.

The reduced density matrix of $A$, after tracing over the remaining system is
\begin{align}
	\tilde \rho_A = \sum_{\bal, \bbe} \tilde \rho_{\bal}^{\bbe} | \bal \rangle \langle \bbe |,
\end{align}
where bold greek symbols denote the set of indices for the subsystem of spins, i.e., all $2^{M_A}$ binary representations of $M_A$ spins: ${\bal}=(\alpha_1,\alpha_2,\dots,\alpha_{M_A})$ with $\alpha_k \in \{0,1\}$. The diagonal elements of $\tilde \rho_A$ can be easily measured. They are the probabilities for finding the spin combination $\bal$, $p_{\bal}$. The off-diagonal elements are more challenging but reduce to the measurement of spin correlations. 

For our experimental situation, we can make use of the fact that for any time-evolved state $\ket{\psi(t)}=\sum_{\bal} c_{\bal}(t) \ket{\bal}$,
\begin{align}
	c_{\bal}(t)=0 \Leftrightarrow \bigoplus_k^M \alpha_k = 1,
\end{align}
where $\bigoplus$ denotes the sum modulo $2$. This result is easily verified by the fact that the matrix elements of any power of Hamiltonian~\eqref{eq:ising_ham} $\bra{\bal} \hat H^n \ket{\bbe}=0$ for $\bigoplus_k^M \alpha_k\neq\bigoplus_k^M \beta_k$. Thus, since we start in the state with $c_{\bal}(t=0)=\delta_{{\bal},{\bf 0}}$,  the time-evolution operator $\exp(-it\hat H)=\sum_{n=0} {(-it\hat H)^n}/{n!}$ can only produce states with nonzero coefficients $c_{\bal}(t)$ for which $\bigoplus_k^M \alpha_k = 0$.
Thus, half the elements of any reduced density matrix calculated from the time-evolved state $\ket{\psi(t)}$ will always remain zero. The remaining spin-spin correlations consist only of $\sigma_x$ and $\sigma_y$ terms. 

We illustrate this here for the example of a subsystem of a single spin ($l=1$) and two spins ($l=2$), and show how to reconstruct the corresponding density matrix. In the case of a single spin, the density matrix will be diagonal:
\begin{align}	
	\tilde \rho_1 = p_0 \ket{0}\bra{0} + p_1\ket{1}\bra{1}
	.
\end{align}
Here, the off-diagonal part completely vanishes since trivially, for all $\tilde \rho_\alpha^\beta$ with $\alpha\neq\beta$, $\alpha \oplus \beta=1$. For a block of two spins, the density matrix becomes
\begin{align}	
	\tilde \rho_2 =
\left(\begin{array}{cccc}p_{00} & 0 & 0 & p_{00}^{11} \\0 & p_{01} & p_{01}^{10} & 0 \\0 & \left(p_{01}^{10}\right)^* & p_{10} & 0 \\\left(p_{00}^{11}\right)^* & 0 & 0 & p_{11}\end{array}\right)
	.
\end{align}
The six density-matrix values are all that have to be measured. 
To experimentally obtain the off-diagonal elements, one has to measure the real and imaginary parts of $\rho_{00}^{11}$ and $\rho_{01}^{10}$. This can be done by expanding those elements into spin-spin correlations via
\begin{align}	
	{\rm Re}(\rho_{00}^{11})
	&= \left(\hat \sigma^x \otimes  \hat \sigma^x - \hat \sigma^y \otimes  \hat \sigma^y \right)/4, \\
	{\rm Im}(\rho_{00}^{11})
	&=
	 \left(\hat \sigma^x \otimes  \hat \sigma^y + \hat \sigma^y \otimes  \hat \sigma^x \right)/4,\\
	 {\rm Re}(\rho_{01}^{10})
	&= \left(\hat \sigma^x \otimes  \hat \sigma^x + \hat \sigma^y \otimes  \hat \sigma^y \right)/4, \\
	{\rm Im}(\rho_{00}^{11})
	&=
	 \left(\hat \sigma^y \otimes  \hat \sigma^x - \hat \sigma^x \otimes  \hat \sigma^y \right)/4
	 ,
\end{align}
which means that four spin correlations have to be measured. Thus, for the whole density matrix $\tilde \rho_2$, a total of eight expectation values have to be experimentally determined. 
Note that this is a simplified version of full quantum-state tomography for two qubits \cite{roos_bell_2004}; however, a full state tomography, i.e., a measurement of all density matrix elements might be useful to quantify the deviations of the other elements from zero and therefore get a measure for experimental deviations. 

From the density matrices, one can directly extract the corresponding von Neumann entropy. Note, however, that one can also calculate Renyi entropies of arbitrary order $n$. These entropies are defined as $S^{[n]}(\rho)=\log_2[{\rm tr}(\tilde \rho)]/(1-n)$ and can serve as lower bounds to the von Neumann entropy.  In an analogous fashion, one can use these entropies to define a mutual information

A generalization of the density-matrix measurement to larger blocks is straightforward. In particular, for a block length of $l$ sites, one has to measure a total amount of $[2^{2l-2}+2^{(l-1)}]$ density-matrix values. For example, to clearly observe a linear entanglement increase for $\alpha =2$, one would have to simulate the system until a time of $\sim2/\bar J$ [cf.~Fig.~\ref{fig:crossover}(a)]. Before finite-size effects of the block size play a role, one therefore has to consider a block with $l=8$. While this seems to be achievable in current experiments (full state tomography for systems consisting of eight ions have been reported in \cite{haffner_scalable_2005}), we emphasize that ultimately the direct measurement of entanglement entropies of larger blocks becomes essentially impossible. This result is due to the fact that the number of correlation functions that have to be measured experimentally increases exponentially with the block size $l$. However, it is important to note that the mutual information can always be extracted by only using blocks of $l=1$ and $l=2$.

Alternatively to measuring the mutual information, the problem of measuring entanglement entropies of large blocks  could be overcome by using recently proposed measurement schemes, which rely on the preparation of identical copies of the same state  \cite{horodecki_method_2002,cardy_measuring_2011}.  In this case, either Rabi oscillations of a quantum switch (coupled to the copies) \cite{abanin_measuring_2012} or ``beam-splitter'' operations between those copies and repeated measurements of the spin configurations \cite{daley_measuring_2012} could be used for an experimental estimation of Renyi entropies. Multiple copies could, for example, be realized by performing the quench identically for cotrapped strings in microtraps. Alternatively, one could use a single, large string and hiding pulses to effectively realize two copies.

\section{Conclusion \& Outlook} \label{sec:conclusion}

We have studied the dynamical evolution of entanglement in quench experiments in ion traps. We found that a regime of linear half-chain von Neumann entropy growth is present even for relatively long-range interactions with interaction decay exponents of $\alpha \gtrsim 1$. The growth rate is sensitive to the underlying low-energy spectrum and notably largest at the point of the quantum phase transition, which varies with changing $\alpha$. For longer-range interactions, we find a regime of logarithmic entropy growth ($\alpha \lesssim 1$), and for infinite-range interactions ($\alpha=0$), the entropy remains bound by a small constant and oscillates rapidly. We showed that mutual information between distant parts of the system can be experimentally measured and used to distinguish the different regimes.

The entanglement entropy growth behavior has important implications for t-DMRG/MPS/MPO algorithms on classical computers since it underlies the complexity of these algorithms. For a specific experimental example, we have shown that this regime is in reach for an experimental quantum simulator, which means that we can find a regime in these systems that fulfills a necessary condition realization of a quantum simulator regime where state-of-the-art numerical simulations on classical computers can become inefficient. While this idea provides a general strong motivation for experiments in this regime, we emphasize, that in future work also the study of dynamics of different types of, e.g., multipartite entanglement as a resource for specific quantum-information applications could be interesting. 

We emphasize that ion-trap experiments are not the only experimental realizations in which the entanglement dynamics studied here could be observed. For example, exactly the same spin model \cite{schachenmayer_dynamical_2010} and also more complicated models with long-range interactions can be realized in systems with polar molecules \cite{gorshkov_tunable_2011,chotia_long-lived_2012} or Rydberg atoms \cite{lesanovsky_many-body_2011,weimer_rydberg_2010,viteau_rydberg_2011, schaus_observation_2012} in optical lattices. These systems would have the disadvantage that the decay exponent is not directly tunable; however, they could have the advantage that half-chain entropies might be easier to measure directly, by employing schemes that rely on the preparation of multiple copies in an optical lattice \cite{daley_measuring_2012}.

In the final stages of preparing our manuscript, we became aware of some related work \cite{hauke_spread_2013}, in which a {\em local} quench in an Ising model with algebraically decaying interactions is studied, together with the resulting spread of quasiparticles. Though our quenches are qualitatively different, and large-scale-entanglement growth cannot be observed in a local quench, Ref.~\cite{hauke_spread_2013} has interesting parallels with what we observe here, and it identifies similar parameter regimes for dynamics with long-range and short-range interactions. 
 
\emph{Acknowledgements -} We thank Rainer Blatt, Ignacio Cirac, Stephan Langer, Hannes Pichler, and Peter Zoller for stimulating discussions. Work in Pittsburgh is supported by NSF Grant No. PHY-1148957, with code development supported by AFOSR Project No. FA9550-12-1-0057. Computational resources were provided by the Center for Simulation and Modeling at the University of Pittsburgh. This work was partially supported by the Austrian Research Fund under Project No. P25354-N20.

\begin{appendix}

\section{Numerical  simulations and entanglement growth}
\label{app:numerics}

In this appendix, we will give more details about the connection of the entanglement growth behavior and the numerical algorithms we use.

\subsection{MPS simulations vs entanglement growth}

There is an interesting connection between the time-dependent growth of the half-chain von Neumann entropy and the possibility to simulate dynamics on classical computers. The dimension of the reduced density matrix for half of the system is $\dim(\rho_L)=2^{M/2}$
and thus grows exponentially as a function of the system size. This
exponential Hilbert space growth is the reason why exact numerical
simulation becomes, in practice, impossible for large system sizes. The von Neumann entropy is defined as
 \begin{align} S_{\rm vN}\equiv S(\rho_L) \equiv -{\rm tr}
 \left( \rho_L \log_2 \rho_L \right) = -\sum_{\alpha}^{\chi_L}
 \lambda_\alpha \log_2(\lambda_\alpha),
 \end{align} 
  where in the second
 equation we defined the eigenvalues of $\rho_L$ as $\lambda_\alpha$ and
 introduced the number of nonzero eigenvalues,
 $\chi_L\leq\dim(\rho_L)$. $\chi_L$ is called the Schmidt rank and can
 be considered as an entanglement measure itself. 
Also note that the maximum possible Schmidt rank grows
exponentially as a function of $M$, and correspondingly, the maximum
possible von Neumann entropy grows {\rm linearly} as $S_{\rm vN}^{[{\rm
max}]}=M/2$. However, the Schmidt rank for states as they occur in typical experiments can be much
smaller than the dimension of $\rho_L$. The approximation, which is made
in numerical DMRG/MPS algorithms (see below), consists therefore of
truncating the Schmidt rank (for all possible bipartite splittings) at a
maximum value, which is called the bond dimension $D$, thus effectively
limiting the von Neumann entropy to $S_{\rm vN}^{[{\rm
MPS}]}\leq\log_2(D)$. The error made is called the truncated weight,
$\epsilon_D=\sum_{i=D+1}^{\chi_L} \lambda_\alpha$.

The truncation made in typical DMRG simulations is very significant. For
example, considering a string of $50$ spins, the dimension of $\rho_L$
is given by $2^{25}$. If one represents a state of this system by a MPS
with a bond dimension of $D_{M/2}=1024$, the size of the effective
Hilbert space for one-half of the system is still only around $0.003\%$ of the
full Hilbert space. However, it is quite remarkable that this
approximation is, in many cases, quasi exact. The reason for this is that
most low-energy states of physical systems in nature turn out to be, in fact,
very slightly entangled. If, for example, in Hamiltonian
\eqref{eq:ising_ham} we restrict the interaction to nearest neighbors
(standard transverse Ising model), then it can be proven that for the
critical model, i.e., for $J=B$, where for an infinite system the energy
gap from the ground state to excited states disappears, the entanglement
entropy scales with the block length as $S_L^{[\rm GS]}\sim \log_2(L)$.
If the system is gapped, the entanglement entropy scales as $S_L\sim
{\rm const.}$, i.e., it obeys an ``area law''
\cite{amico_entanglement_2008,eisert_colloquium:_2010}. Thus, in the
worst case (critical model), the bond dimension only has to grow linearly
with the system size, and therefore ground-states can be easily
calculated up to a quasi exact precision with DMRG/MPS for systems of
hundreds of spins.

It is obvious that, therefore, in a time evolution simulation, whether the
simulation of the system over long times with t-DMRG methods is, in practice,
possible or not depends on how fast the entanglement grows. If in our
quench experiment with ions, the von Neumann entropy grows linearly
as a function of time, in order to keep $\epsilon_D$ small, $D$ has to
grow exponentially as a function of time. The computational resources to store a MPS thus grows exponentially with the time the system is to be simulated, which becomes prohibitively expensive for
large system sizes. 

\subsection{Exact diagonalization}

Despite the exponential growth of the Hilbert space, quantum systems of moderate size can still be diagonalized exactly, simply by exploiting the sparseness of typical Hamiltonians. For example, for $20$ spins, the full Hamiltonian is a $2^{20}\times 2^{20}$ matrix, which is too large to even store in the memory of current computer hardware. However the amount of nonzero elements of Hamiltonian  \eqref{eq:ising_ham} is only of $\mathcal{O}(2^{20})$, even for a full interaction matrix. Therefore, one can use Krylov subspace projection techniques  \cite{sidje_expokit:_1998} to evaluate the matrix exponential of the Hamiltonian matrix, as well as semi-implicit first-order methods to propagate a state vector for systems of $20$ spins in time.

\subsection{MPO/MPS algorithms}

For larger systems, one has to use an alternative approximate spin-representation in form of a matrix product state (MPS) \cite{vidal_efficient_2004,
  white_real-time_2004, daley_time-dependent_2004,schollwock_density-matrix_2011,verstraete_matrix_2008}. A MPS is defined as the decomposition of the complex amplitudes of the full
quantum state of a lattice system, $\ket{\psi} = \sum_{\{i_k\}} c_{i_1,i_2,\dots,i_M} \ket{i_1} \ket{i_2} \dots \ket{i_M}$ ($M$
sites with local basis states $\{\ket{i_k}\}$) into a matrix product.
Specifically, we define a MPS in its canonical form as 
\begin{align}
\label{eq:canmps_definition} c_{i_1,i_2,\dots,i_M} \equiv A_{i_1}^{[1]}
S^{[1]} A_{i_2}^{[2]} S^{[2]} \dots S^{[M-1]} A_{i_M}^{[M]} .
\end{align} 
Here, the $A^{[k]}_{i_k}$ are complex unitary $D_{i-1}\times
D_i$ matrices in an effective basis, and for open boundary conditions,
$D_0=D_M=1$. The $S^{[i]}$ are real diagonal $D_{i}\times D_{i}$
matrices with unit norm, $S^{[i]\,\dag}S^{[i]}=1$. Any arbitrary state
can be brought into the form \eqref{eq:canmps_definition} by making use
of subsequent singular-value decompositions of the $M$--dimensional
tensor $c_{i_1,i_2,\dots,i_M}$ (see,e.g., \cite{schollwock_density-matrix_2011}). In general, the sizes of the
matrix dimensions that are required to represent a certain state exactly
are given by the Schmidt rank $\chi_i$ for the bipartite splitting
between sites $i$ and $i+1$. Limiting all matrix dimensions by the
bond-dimension $D$ limits the maximum allowed von Neumann entropy to
$S_{\rm vN}^{[{\rm MPS}]}\leq\log_2(D)$. Note that in
the case of $D=1$, $S_{\rm vN}^{[{\rm MPS}]}=0$, and it is readily seen
that the MPS reduces to a nonentangled product state.

We can study systems with long-range interactions numerically by
making use of matrix product operators (MPOs), which are as the MPSs, a
decomposition of the now real $4^M$-dimensional operator tensor of
the full Hamiltonian $H = \sum_{\{i_k\},\{j_k\}}
o_{i_1,i_2,\dots,i_M}^{j_1,j_2,\dots,j_M} \ket{i_1}\ket{i_2} \dots
\ket{i_M} \bra{j_1}\bra{j_2}\dots \bra{j_M}$ into a matrix
product. The long-range interaction Hamiltonian \eqref{eq:ising_ham}
with a decaying interaction $J_{ij}=\bar J /|i-j|^{\alpha}$ can be, up
to a very good approximation, implemented as MPO with relatively small
bond-dimension, which can be achieved by expanding the power-law decay
function into a sum of exponentials
\cite{crosswhite_applying_2008,pirvu_matrix_2010,frowis_tensor_2010}.
With the Hamiltonian in MPO form it is then possible to implement
time-evolution algorithms, using e.g., a Runge-Kutta-type evolution
scheme \cite{garcia-ripoll_time_2006}. Alternatively, one can also use
the original adaptive t-DMRG methods \cite{vidal_efficient_2004,
  white_real-time_2004, daley_time-dependent_2004}  and introduce
swap-gates, which interchange indices in a MPS. This has the advantage
that arbitrary interaction matrices $J_{ij}$ can be implemented.

To calculate ground states, one can either evolve a MPS in negative imaginary time, or one can construct the local representations of the Hamiltonian expectation value by leaving the indices on a particular site open and contracting the remaining tensor network (see
e.g., \cite{schollwock_density-matrix_2011}). We then use a local iterative Lanczos solver to find the local MPS matrix, which minimizes the corresponding energy. By sweeping through the system from site to site, this is a very efficient method to find the overall MPS ground state for large systems.

We checked the validity of all our results by comparing different methods, and we confirm the convergence in the bond dimension by running multiple calculations with increasingly large $D$.

\section{Details on analytical calculations}
\label{app:analytics}

\subsection{Quasiparticle contribution to the half-chain entropy growth}

Here, we give more details for the quasiparticle picture for the nearest-neighbor Ising model.
Following \cite{pfeuty_one-dimensional_1970}
the nearest neighbor transverse Ising Hamiltonian \eqref{eq:ising_nn_ham}
 can be rewritten in terms of local spin-lowering and spin-raising operators, $\sigma_j^\pm\equiv (\sigma_j^x\pm{\rm i} \sigma_j^y)/2$. With a Jordan-Wigner transformation, these operators can be mapped to anti-commuting quasiparticles via $c_i = \left[{\exp}\left({{\rm i} \pi \sum_{j=1}^{i-1} \sigma_j^+ \sigma_j^-} \right)\right] \sigma^-_i = \prod_{j=1}^{i-1} (1-2 \sigma_j^+ \sigma_j^-) \sigma_i^-$. We thus end up with a fermionic Hamiltonian
\begin{align}
 H=J \sum_{i} (c_i^\dag - c_i) (c_{i+1}^\dag + c_{i+1}) + B \sum_i (c_i^\dag c_i -c_i c_i^\dag).
\end{align}
 Assuming translational invariance and expanding the quasiparticle operators into plane waves, the Hamiltonian in quasimomentum space becomes
\begin{align}
H=\sum_{q>0} 
\begin{pmatrix}
c_q^\dag & c_{-q} 
\end{pmatrix}
\begin{pmatrix}
\tilde \epsilon_q & 2 iJ \sin(qa) \\
-2 iJ \sin(qa) & -\tilde\epsilon_q \\
\end{pmatrix}
\begin{pmatrix}
c_q \\ 
c_{-q}^\dag
\end{pmatrix}
\end{align}
with anticommuting fermions $c_q$, $\{c_q,c_q'\}=\delta_{q,q'}$ and $\tilde \epsilon_q=2J\cos(qa)+2B$. For $M$ spins, the quasi momenta are given by $q=n 2 \pi/(aM)$, where $a$ is the separation of the spins, and $n=-M/2, \dots, M/2 -1$. The Hamiltonian can be diagonalized using a unitary (Bogoliubov) transformation, where the new fermionic quasiparticles $\gamma$ are given by $(\gamma_q^\dag , \gamma_{-q}) = U_B (c_q^\dag, c_{-q})$. Performing this transformation leads to the diagonal model
\begin{align}
H=\sum_{q} {\epsilon}_q \left(\gamma_q^\dag \gamma_q - \frac{1}{2} \right).
\end{align}
where the dispersion relation of the new particles is given by $\epsilon_q= 2
\sqrt{(J-B)^2+4JB \sin^2(qa/2)}$.

\begin{figure}[tb] 
\begin{center}
\includegraphics[width=0.48\textwidth]{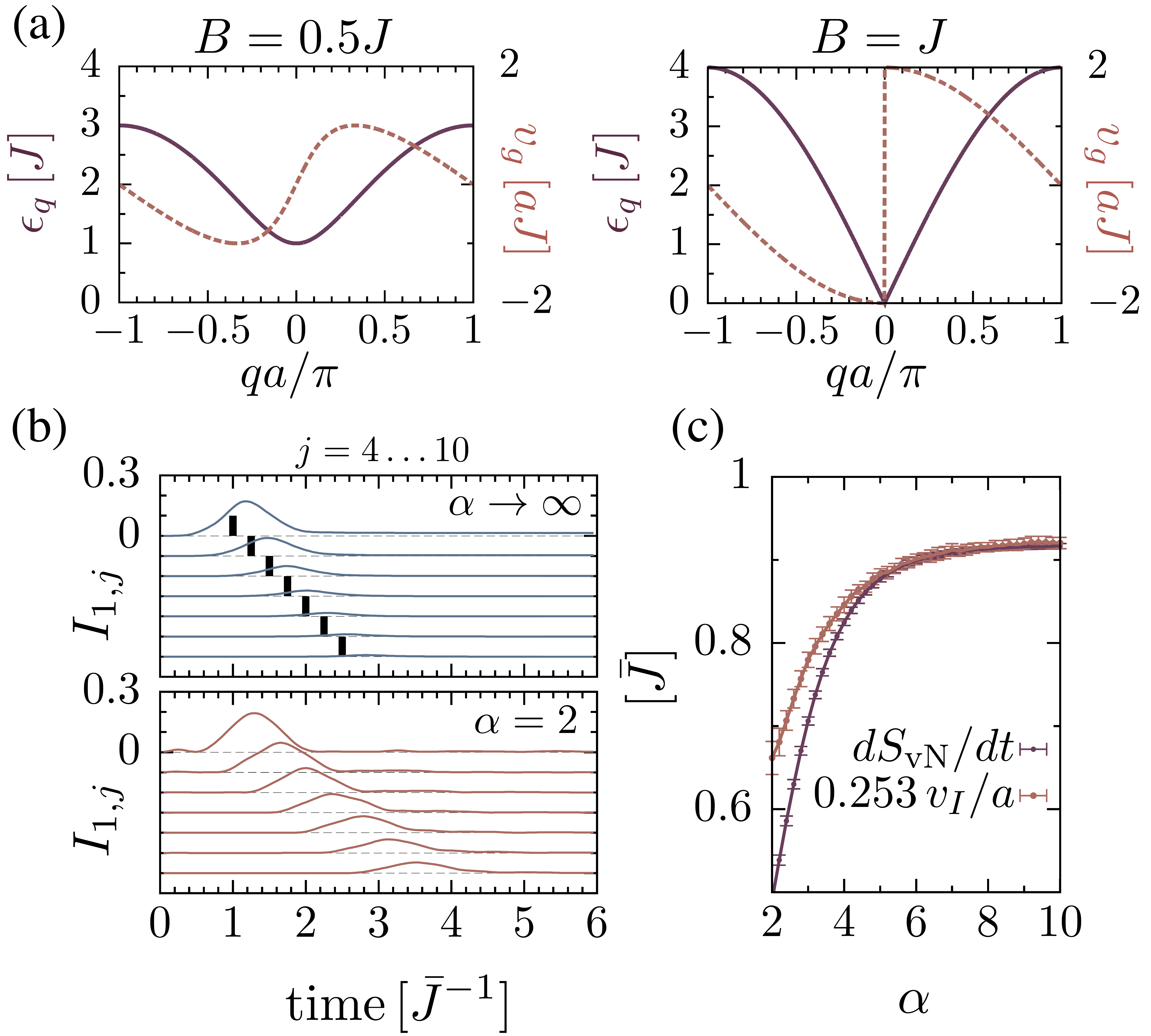} 
\end{center}
\caption{(a) Quasiparticle dispersion relations for the nearest-neighbor Ising Hamiltonian for $B=0$ and $B=J$, $\epsilon_q$. The corresponding groupvelocities $v_g(q)/a=d\epsilon_q/d(qa)$ are shown as dashed lines (right axis). (b) The time evolution of the mutual information $I_{1,j}$ between the leftmost spin and $j=4\dots 10$ (ED, $B=1$, $M=20$; different $I_{1,j}$ offset by $-0.1$ for different $j$ for better visibility). The upper panel is for $\alpha\rightarrow\infty$ (nearest-neighbor interactions), the lower panel for $\alpha=2$. The vertical black bars indicate the analytical result of the entangled quasiparticle pair arrival, calculated from the group velocity. (c) The half-chain entropy growth rates (linear fit $1 < t\bar J < 3$,  ED, $B=1$, $M=20$) compared to the  effective velocity of the entangling ``quasiparticle wave'' $v_I$ (see text). We extract $v_I$ from a linear fit to the position of the mutual information peaks in panel (c).  \label{fig:sp_spectrum}}
\end{figure}

Example dispersion relations and corresponding group velocities $v_g(q)/a=d\epsilon_q/d(qa)$ are shown in Fig.~\ref{fig:sp_spectrum}(a). At the critical point, the gap closes, and the dispersion relation is linear around $q=0$. In this regime, the particles with the Lieb-Robinson velocity $|v_R|$ are found for $q\sim 0$, whereas for $B\neq J$, the fastest particles shift to larger $|q|$. For $B \gg J$ and $B \ll J$, the fastest quasiparticles are found at  $q\sim \pm \pi/2$.

We can analyze how the propagation of free quasiparticle pairs contributes to the entanglement growth by looking at the mutual information between distant spins. In Fig.~\ref{fig:sp_spectrum}(b) we plot the mutual information $I_{1,j}$ between site $1$ and $j$ while increasing $j=4,\dots,10$ for nearest-neighbor interactions and for $\alpha=2$. We find that in both cases, for two particular spins that are separated by some distance, the mutual information remains nearly zero for a long time, until it suddenly peaks at a time corresponding to the arrival of a quasiparticle pair originally produced on a site between the two spins, which then entangles the two spins. After the quasiparticles pass, the mutual information remains at a value slightly greater than zero (barely visible). We find that the time of the arrival of the wave at site $j$ is consistent with quasiparticles moving at the Lieb-Robinson velocity for nearest neighbor interactions. Since the two sites become entangled once a quasiparticle that has been created in the middle of the two sites first arrives at both spins, this time is given by $t_j=(j-2)/2v_R=(j-2)/2a\bar J$ and is shown as black bars in Fig.~\ref{fig:sp_spectrum}(b). In the lower panel, we find that even for $\alpha=2$, despite the rather long-range interactions, the characteristic behavior is the same. For longer-range interactions, we find that the wave moves more slowly. Furthermore we note that for $\alpha\rightarrow\infty$ one finds a much more ``diffusive'' behavior, in the sense that the peaks of mutual information broaden and become smaller with distance.

The peaks in Fig.~\ref{fig:sp_spectrum}(b) allow us to extract an effective velocity of the ``wave" of entangling quasiparticles, $v_I$. We do this by fitting a line to the position of the peaks (in time) as a function of the distance $j$ to obtain $1/v_I$. As we show in  Fig.~\ref{fig:sp_spectrum}(c), the rate of the half-block entropy increase (linear fit $1\leq t \bar J \leq 3$) is directly related to $v_I$. We find that for $\alpha \gtrsim 5$, $S_{\rm vN} (t) = \eta v_I t /a$ with $\eta \approx 0.253$. For $\alpha \lesssim 3$, the proportionality constant starts to depend on $\alpha$ and  $\eta < 0.253$. This means that for an increasing range of interactions, both the half-chain entropy growth rate and $v_I$ reduce, and  $v_I$ decreases less strongly. We note that, in general, $\eta$ also depends on $B$.

\subsection{Model for infinite-range interactions in the Dicke-state basis}

Here, we give more details on our calculation for infinite-range interactions, i.e., a
model where each spin interacts with equal strength with all
others. As shown in the main text, in this regime the Hamiltonian for $M$ ions becomes a spin-M/2 model, which is known as the LMG model \cite{vidal_entanglement_2004,wilms_finite-temperature_2012,latorre_entanglement_2005},
\begin{align} 
\hat
H= \frac{J}{2} S_x^2 + B S_z - \frac{J}{2} M
\end{align} 
with a basis given by the Dicke states
\begin{align}
\left|S=\frac{M}{2},m_S=n_1-\frac{M}{2}\right\rangle \equiv \mathcal{S} |\{n_0,n_1 \} \rangle.
\end{align}

We want to calculate the bipartite entanglement for a splitting in the center of the chain. Therefore, we make use of a formula for the Schmidt decomposition of Dicke states \cite{latorre_entanglement_2005}. Specifically, any spin-M/2 Dicke state can be rewritten as a sum over product states of two spin-M/4 Dicke states, for the left and right halves of the system, respectively. For $n_1$ spins up, this decomposition can be written as
\begin{align}
\label{eq:dicke_sd}
&\left| \frac{M}{2},n_1-\frac{M}{2} \right\rangle =\sum_{l=0}^{M/2} \sqrt{p_l(n_1)} \nonumber\\ 
& \qquad \times 
\left| \frac{M}{4},l-\frac{M}{4} \right\rangle_L \left| \frac{M}{4},n_1-l-\frac{M}{4}\right\rangle_R.
\end{align}
The $p_l$ can be found using combinatorical arguments,
\begin{align}
p_l(n_1) &= \frac{ {M/2\choose l} {M/2\choose n_1-l} }{ {M \choose n_1}}
.
\end{align}
The von Neumann entropy of half the chain can be trivially extracted from the $p_l$ since they are simply the eigenvalues of the reduced density matrix for half the chain. Therefore, $S_{\rm vN}=-\sum_l p_l \log_2(p_l)$. 

In our quench experiment, initially the ion chain is in the Dicke state $\ket{M/2,-M/2}$, and the subsequent time evolution will rotate the state vector in the Dicke manifold and therefore prepare a time-dependent superposition 
\begin{align}
\label{eq:dicke_sup}
|\psi(t)\rangle=\sum_m c_m(t) |{M/2,m}\rangle
\end{align}
 Numerically determining the $c_m (t)$ is easily 
achieved for large systems, since the dimension of the Hilbert space is only given by $M+1$. However, when inserting the formula \eqref{eq:dicke_sd} into \eqref{eq:dicke_sup}, the resulting decomposition is not a proper Schmidt decomposition anymore since the``Schmidt values'' can be complex now. Therefore, one has to construct the full reduced density matrix by tracing over one-half of the system,
\begin{align}
\rho_L=\sum_{m_r} \left\langle \frac{M}{4}, m_r \right|
\psi(t) \bigg\rangle\bigg\langle \psi(t)
\left| \frac{M}{4}, m_r \right\rangle
\end{align}
Performing the partial trace, we find
\begin{align}
\label{eq:time_red_dm}
&\rho_L (t) = \sum_l \sum_{\tilde m,m} 
c_{\tilde m}^*(t) \sqrt{p_l(\tilde m)} 
c_{\tilde m}(t) \sqrt{p_l(m)} 
\nonumber\\ 
&\qquad \times 
 \left| \frac{M}{4}, \tilde m +\frac{M}{4} -l \right\rangle \left\langle \frac{M}{4}, m +\frac{M}{4} -l \right|_L
 .
\end{align}
This matrix can then be easily diagonalized numerically, which allows us to time dependently calculate the half-chain entropy.

For the same reason as discussed in Sec.~\ref{sec:measurement}, we find that half the coefficients $c_m(t)$ will always remain zero. The Hamiltonian only couples terms with $m \leftrightarrow m\pm 2$. Thus the amount of nonzero eigenvalues of the matrix \eqref{eq:time_red_dm} is limited by $M/4+1$ (for even $M$). This gives the upper bound of the entropy, $S_{\rm vN}\leq \log_2(M/4+1)$.

\section{Realistic experiment with 20 ions}
\label{app:exp}

In Sec.~\ref{sec:experiment} the entanglement growth is analyzed in experimentally realistic situations. Here, we give details on the exact experimental parameters considered. In each case, we consider a string of 20 ions in a 3D harmonic trap with highest transverse motional frequency $\omega_t{=}2\pi{\times}4.9$ MHz  and variable lowest axial frequency $0.1{<}\omega_z/2\pi{<}0.5$ MHz. A state-dependent driving force is simulated at a fixed detuning $\delta{=}{+}2\pi{\times}80$ kHz from $\omega_t$. These three parameters are sufficient to completely determine the form of the interaction matrix \cite{PhysRevLett.103.120502, 1367-2630-14-9-095024}. 

In order to vary the interaction strength we choose to vary $\omega_z$, which has the effect of changing how closely spaced the transverse mode are. For the cases shown in Figs.~\ref{fig:real_experiment}(a)--(c), we choose $\omega_z{=}2\pi{\times} 0.45, 2\pi{\times} 0.25,$ and $ 2\pi{\times} 0.1$ MHz, respectively. Alternatively, one could fix $\omega_z$  and change the detuning $\delta$.  

The absolute value of the interaction strength $\bar{J}$ is determined by the specific choice of driving-force mechanism, driving strength, ionic species, and electronic transition used to encode the spin. We consider an optical transition, at 729\,nm, between the $S_{1/2}, m{=}1/2$, ground and metastable $D_{5/2}, m{=}3/2$, states in $^{40}$Ca$^{+}$. Copropagating bichromatic laser fields at 729\,nm drive the interaction, with a Rabi frequency of $\Omega{=}2\pi \times 0.5$ MHz (if put on resonance with the electronic spin). For the highest transverse mode cooled to the ground state, the coupling strength on the upper vibrational sideband of the transition is therefore $\eta_t\Omega{=}2\pi\times 22$~kHz, where $\eta_t{=}0.044$ is the Lamb-Dicke parameter. In the case $\delta\gg\eta_t\Omega$, the assumption that the phonon states can be adiabatically eliminated holds. In our case $\delta \approx 4 \eta_t\Omega$. This approximation could be improved at the expense of a slower overall interaction strength $\bar{J}$.  

For the cases shown in Fig.~\ref{fig:real_experiment}(a)--(c), we find overall spin-spin coupling rates $\bar{J}/2\pi{=} 2\pi \times 0.5, 2\pi \times0.4$ and $2\pi \times 0.3$ kHz. These rates correspond to quantum dynamics observable on time scales of a few ms, which compares favorably with typical decoherence times of several 10's of ms that are typically achieved in these experiments.

\end{appendix}

\bibliographystyle{apsrev4-1-custom}
\bibliography{ion}

\begin{thebibliography}{72}%
\makeatletter
\providecommand \@ifxundefined [1]{%
 \@ifx{#1\undefined}
}%
\providecommand \@ifnum [1]{%
 \ifnum #1\expandafter \@firstoftwo
 \else \expandafter \@secondoftwo
 \fi
}%
\providecommand \@ifx [1]{%
 \ifx #1\expandafter \@firstoftwo
 \else \expandafter \@secondoftwo
 \fi
}%
\providecommand \natexlab [1]{#1}%
\providecommand \enquote  [1]{``#1''}%
\providecommand \bibnamefont  [1]{#1}%
\providecommand \bibfnamefont [1]{#1}%
\providecommand \citenamefont [1]{#1}%
\providecommand \href@noop [0]{\@secondoftwo}%
\providecommand \href [0]{\begingroup \@sanitize@url \@href}%
\providecommand \@href[1]{\@@startlink{#1}\@@href}%
\providecommand \@@href[1]{\endgroup#1\@@endlink}%
\providecommand \@sanitize@url [0]{\catcode `\\12\catcode `\$12\catcode
  `\&12\catcode `\#12\catcode `\^12\catcode `\_12\catcode `\%12\relax}%
\providecommand \@@startlink[1]{}%
\providecommand \@@endlink[0]{}%
\providecommand \url  [0]{\begingroup\@sanitize@url \@url }%
\providecommand \@url [1]{\endgroup\@href {#1}{\urlprefix }}%
\providecommand \urlprefix  [0]{URL }%
\providecommand \Eprint [0]{\href }%
\providecommand \doibase [0]{http://dx.doi.org/}%
\providecommand \selectlanguage [0]{\@gobble}%
\providecommand \bibinfo  [0]{\@secondoftwo}%
\providecommand \bibfield  [0]{\@secondoftwo}%
\providecommand \translation [1]{[#1]}%
\providecommand \BibitemOpen [0]{}%
\providecommand \bibitemStop [0]{}%
\providecommand \bibitemNoStop [0]{.\EOS\space}%
\providecommand \EOS [0]{\spacefactor3000\relax}%
\providecommand \BibitemShut  [1]{\csname bibitem#1\endcsname}%
\let\auto@bib@innerbib\@empty
\bibitem [{\citenamefont {Bloch}\ \emph {et~al.}(2012)\citenamefont {Bloch},
  \citenamefont {Dalibard},\ and\ \citenamefont
  {Nascimb\'ene}}]{bloch_quantum_2012}%
  \BibitemOpen
  \bibfield  {author} {\bibinfo {author} {\bibfnamefont {I.}~\bibnamefont
  {Bloch}}, \bibinfo {author} {\bibfnamefont {J.}~\bibnamefont {Dalibard}}, \
  and\ \bibinfo {author} {\bibfnamefont {S.}~\bibnamefont {Nascimb\'ene}},\
  }\bibfield  {title} {{\selectlanguage {english}\enquote {\bibinfo {title}
  {Quantum simulations with ultracold quantum gases},}\ }}\href {\doibase
  10.1038/nphys2259} {\bibfield  {journal} {\bibinfo  {journal} {Nat. Phys.}\
  }\textbf {\bibinfo {volume} {8}},\ \bibinfo {pages} {267--276} (\bibinfo
  {year} {2012})}\BibitemShut {NoStop}%
\bibitem [{\citenamefont {Blatt}\ and\ \citenamefont
  {Roos}(2012)}]{blatt_quantum_2012}%
  \BibitemOpen
  \bibfield  {author} {\bibinfo {author} {\bibfnamefont {R.}~\bibnamefont
  {Blatt}}\ and\ \bibinfo {author} {\bibfnamefont {C.~F.}\ \bibnamefont
  {Roos}},\ }\bibfield  {title} {{\selectlanguage {english}\enquote {\bibinfo
  {title} {Quantum simulations with trapped ions},}\ }}\href {\doibase
  10.1038/nphys2252} {\bibfield  {journal} {\bibinfo  {journal} {Nat. Phys.}\
  }\textbf {\bibinfo {volume} {8}},\ \bibinfo {pages} {277--284} (\bibinfo
  {year} {2012})}\BibitemShut {NoStop}%
\bibitem [{\citenamefont {Aspuru-Guzik}\ and\ \citenamefont
  {Walther}(2012)}]{aspuru-guzik_photonic_2012}%
  \BibitemOpen
  \bibfield  {author} {\bibinfo {author} {\bibfnamefont {A.}~\bibnamefont
  {Aspuru-Guzik}}\ and\ \bibinfo {author} {\bibfnamefont {P.}~\bibnamefont
  {Walther}},\ }\bibfield  {title} {{\selectlanguage {english}\enquote
  {\bibinfo {title} {Photonic quantum simulators},}\ }}\href {\doibase
  10.1038/nphys2253} {\bibfield  {journal} {\bibinfo  {journal} {Nat. Phys.}\
  }\textbf {\bibinfo {volume} {8}},\ \bibinfo {pages} {285--291} (\bibinfo
  {year} {2012})}\BibitemShut {NoStop}%
\bibitem [{\citenamefont {Rigol}\ \emph {et~al.}(2006)\citenamefont {Rigol},
  \citenamefont {Muramatsu},\ and\ \citenamefont
  {Olshanii}}]{rigol_hard-core_2006}%
  \BibitemOpen
  \bibfield  {author} {\bibinfo {author} {\bibfnamefont {M.}~\bibnamefont
  {Rigol}}, \bibinfo {author} {\bibfnamefont {A.}~\bibnamefont {Muramatsu}}, \
  and\ \bibinfo {author} {\bibfnamefont {M.}~\bibnamefont {Olshanii}},\
  }\bibfield  {title} {\enquote {\bibinfo {title} {Hard-core bosons on optical
  superlattices: Dynamics and relaxation in the superfluid and insulating
  regimes},}\ }\href {\doibase 10.1103/PhysRevA.74.053616} {\bibfield
  {journal} {\bibinfo  {journal} {Phys. Rev. A}\ }\textbf {\bibinfo {volume}
  {74}},\ \bibinfo {pages} {053616} (\bibinfo {year} {2006})}\BibitemShut
  {NoStop}%
\bibitem [{\citenamefont {Cazalilla}(2006)}]{cazalilla_effect_2006}%
  \BibitemOpen
  \bibfield  {author} {\bibinfo {author} {\bibfnamefont {M.~A.}\ \bibnamefont
  {Cazalilla}},\ }\bibfield  {title} {\enquote {\bibinfo {title} {Effect of
  suddenly turning on interactions in the {Luttinger} model},}\ }\href
  {\doibase 10.1103/PhysRevLett.97.156403} {\bibfield  {journal} {\bibinfo
  {journal} {Phys. Rev. Lett.}\ }\textbf {\bibinfo {volume} {97}},\ \bibinfo
  {pages} {156403} (\bibinfo {year} {2006})}\BibitemShut {NoStop}%
\bibitem [{\citenamefont {Calabrese}\ and\ \citenamefont
  {Cardy}(2006)}]{calabrese_time_2006}%
  \BibitemOpen
  \bibfield  {author} {\bibinfo {author} {\bibfnamefont {P.}~\bibnamefont
  {Calabrese}}\ and\ \bibinfo {author} {\bibfnamefont {J.}~\bibnamefont
  {Cardy}},\ }\bibfield  {title} {\enquote {\bibinfo {title} {Time dependence
  of correlation functions following a quantum quench},}\ }\href {\doibase
  10.1103/PhysRevLett.96.136801} {\bibfield  {journal} {\bibinfo  {journal}
  {Phys. Rev. Lett.}\ }\textbf {\bibinfo {volume} {96}},\ \bibinfo {pages}
  {136801} (\bibinfo {year} {2006})}\BibitemShut {NoStop}%
\bibitem [{\citenamefont {Kollath}\ \emph {et~al.}(2007)\citenamefont
  {Kollath}, \citenamefont {L\"auchli},\ and\ \citenamefont
  {Altman}}]{kollath_quench_2007}%
  \BibitemOpen
  \bibfield  {author} {\bibinfo {author} {\bibfnamefont {C.}~\bibnamefont
  {Kollath}}, \bibinfo {author} {\bibfnamefont {A.~M.}\ \bibnamefont
  {L\"auchli}}, \ and\ \bibinfo {author} {\bibfnamefont {E.}~\bibnamefont
  {Altman}},\ }\bibfield  {title} {\enquote {\bibinfo {title} {Quench dynamics
  and nonequilibrium phase diagram of the {Bose-Hubbard} model},}\ }\href
  {\doibase 10.1103/PhysRevLett.98.180601} {\bibfield  {journal} {\bibinfo
  {journal} {Phys. Rev. Lett.}\ }\textbf {\bibinfo {volume} {98}},\ \bibinfo
  {pages} {180601} (\bibinfo {year} {2007})}\BibitemShut {NoStop}%
\bibitem [{\citenamefont {L\"auchli}\ and\ \citenamefont
  {Kollath}(2008)}]{lauchli_spreading_2008}%
  \BibitemOpen
  \bibfield  {author} {\bibinfo {author} {\bibfnamefont {A.~M.}\ \bibnamefont
  {L\"auchli}}\ and\ \bibinfo {author} {\bibfnamefont {C.}~\bibnamefont
  {Kollath}},\ }\bibfield  {title} {\enquote {\bibinfo {title} {Spreading of
  correlations and entanglement after a quench in the one-dimensional
  {Bose-Hubbard} model},}\ }\href {\doibase 10.1088/1742-5468/2008/05/P05018}
  {\bibfield  {journal} {\bibinfo  {journal} {J. Stat. Mech. Theor. Exp.}\
  }\textbf {\bibinfo {volume} {2008}},\ \bibinfo {pages} {P05018} (\bibinfo
  {year} {2008})}\BibitemShut {NoStop}%
\bibitem [{\citenamefont {Wichterich}\ and\ \citenamefont
  {Bose}(2009)}]{PhysRevA.79.060302}%
  \BibitemOpen
  \bibfield  {author} {\bibinfo {author} {\bibfnamefont {H.}~\bibnamefont
  {Wichterich}}\ and\ \bibinfo {author} {\bibfnamefont {S.}~\bibnamefont
  {Bose}},\ }\bibfield  {title} {\enquote {\bibinfo {title} {Exploiting quench
  dynamics in spin chains for distant entanglement and quantum
  communication},}\ }\href {\doibase 10.1103/PhysRevA.79.060302} {\bibfield
  {journal} {\bibinfo  {journal} {Phys. Rev. A}\ }\textbf {\bibinfo {volume}
  {79}},\ \bibinfo {pages} {060302} (\bibinfo {year} {2009})}\BibitemShut
  {NoStop}%
\bibitem [{\citenamefont {Worm}\ \emph {et~al.}(2012)\citenamefont {Worm},
  \citenamefont {Sawyer}, \citenamefont {Bollinger},\ and\ \citenamefont
  {Kastner}}]{worm_relaxation_2012}%
  \BibitemOpen
  \bibfield  {author} {\bibinfo {author} {\bibfnamefont {M.~v.~d.}\
  \bibnamefont {Worm}}, \bibinfo {author} {\bibfnamefont {B.~C.}\ \bibnamefont
  {Sawyer}}, \bibinfo {author} {\bibfnamefont {J.~J.}\ \bibnamefont
  {Bollinger}}, \ and\ \bibinfo {author} {\bibfnamefont {M.}~\bibnamefont
  {Kastner}},\ }\bibfield  {title} {\enquote {\bibinfo {title} {Relaxation
  timescales and decay of correlations in a long-range interacting quantum
  simulator},}\ }\href@noop {} {\bibfield  {journal} {\bibinfo  {journal}
  {{arXiv:1209.3697}}\ } (\bibinfo {year} {2012})}\BibitemShut {NoStop}%
\bibitem [{\citenamefont {Foss-Feig}\ \emph {et~al.}(2013)\citenamefont
  {Foss-Feig}, \citenamefont {Hazzard}, \citenamefont {Bollinger},\ and\
  \citenamefont {Rey}}]{foss-feig_nonequilibrium_2013}%
  \BibitemOpen
  \bibfield  {author} {\bibinfo {author} {\bibfnamefont {M.}~\bibnamefont
  {Foss-Feig}}, \bibinfo {author} {\bibfnamefont {K.~R.~A.}\ \bibnamefont
  {Hazzard}}, \bibinfo {author} {\bibfnamefont {J.~J.}\ \bibnamefont
  {Bollinger}}, \ and\ \bibinfo {author} {\bibfnamefont {A.~M.}\ \bibnamefont
  {Rey}},\ }\bibfield  {title} {\enquote {\bibinfo {title} {Nonequilibrium
  dynamics of arbitrary-range {Ising} models with decoherence: An exact
  analytic solution},}\ }\href {\doibase 10.1103/PhysRevA.87.042101} {\bibfield
   {journal} {\bibinfo  {journal} {Phys. Rev. A}\ }\textbf {\bibinfo {volume}
  {87}},\ \bibinfo {pages} {042101} (\bibinfo {year} {2013})}\BibitemShut
  {NoStop}%
\bibitem [{\citenamefont {Hazzard}\ \emph {et~al.}(2013)\citenamefont
  {Hazzard}, \citenamefont {Manmana}, \citenamefont {Foss-Feig},\ and\
  \citenamefont {Rey}}]{hazzard_far--equilibrium_2013}%
  \BibitemOpen
  \bibfield  {author} {\bibinfo {author} {\bibfnamefont {K.~R.~A.}\
  \bibnamefont {Hazzard}}, \bibinfo {author} {\bibfnamefont {S.~R.}\
  \bibnamefont {Manmana}}, \bibinfo {author} {\bibfnamefont {M.}~\bibnamefont
  {Foss-Feig}}, \ and\ \bibinfo {author} {\bibfnamefont {A.~M.}\ \bibnamefont
  {Rey}},\ }\bibfield  {title} {\enquote {\bibinfo {title}
  {Far-from-equilibrium quantum magnetism with ultracold polar molecules},}\
  }\href {\doibase 10.1103/PhysRevLett.110.075301} {\bibfield  {journal}
  {\bibinfo  {journal} {Phys. Rev. Lett.}\ }\textbf {\bibinfo {volume} {110}},\
  \bibinfo {pages} {075301} (\bibinfo {year} {2013})}\BibitemShut {NoStop}%
\bibitem [{\citenamefont {Sodano}\ \emph {et~al.}(2010)\citenamefont {Sodano},
  \citenamefont {Bayat},\ and\ \citenamefont {Bose}}]{PhysRevB.81.100412}%
  \BibitemOpen
  \bibfield  {author} {\bibinfo {author} {\bibfnamefont {P.}~\bibnamefont
  {Sodano}}, \bibinfo {author} {\bibfnamefont {A.}~\bibnamefont {Bayat}}, \
  and\ \bibinfo {author} {\bibfnamefont {S.}~\bibnamefont {Bose}},\ }\bibfield
  {title} {\enquote {\bibinfo {title} {Kondo cloud mediated long-range
  entanglement after local quench in a spin chain},}\ }\href {\doibase
  10.1103/PhysRevB.81.100412} {\bibfield  {journal} {\bibinfo  {journal} {Phys.
  Rev. B}\ }\textbf {\bibinfo {volume} {81}},\ \bibinfo {pages} {100412}
  (\bibinfo {year} {2010})}\BibitemShut {NoStop}%
\bibitem [{\citenamefont {Cheneau}\ \emph {et~al.}(2012)\citenamefont
  {Cheneau}, \citenamefont {Barmettler}, \citenamefont {Poletti}, \citenamefont
  {Endres}, \citenamefont {Schau{\ss}}, \citenamefont {Fukuhara}, \citenamefont
  {Gross}, \citenamefont {Bloch}, \citenamefont {Kollath},\ and\ \citenamefont
  {Kuhr}}]{cheneau_light-cone-like_2012}%
  \BibitemOpen
  \bibfield  {author} {\bibinfo {author} {\bibfnamefont {M.}~\bibnamefont
  {Cheneau}}, \bibinfo {author} {\bibfnamefont {P.}~\bibnamefont {Barmettler}},
  \bibinfo {author} {\bibfnamefont {D.}~\bibnamefont {Poletti}}, \bibinfo
  {author} {\bibfnamefont {M.}~\bibnamefont {Endres}}, \bibinfo {author}
  {\bibfnamefont {P.}~\bibnamefont {Schau{\ss}}}, \bibinfo {author}
  {\bibfnamefont {T.}~\bibnamefont {Fukuhara}}, \bibinfo {author}
  {\bibfnamefont {C.}~\bibnamefont {Gross}}, \bibinfo {author} {\bibfnamefont
  {I.}~\bibnamefont {Bloch}}, \bibinfo {author} {\bibfnamefont
  {C.}~\bibnamefont {Kollath}}, \ and\ \bibinfo {author} {\bibfnamefont
  {S.}~\bibnamefont {Kuhr}},\ }\bibfield  {title} {{\selectlanguage
  {english}\enquote {\bibinfo {title} {Light-cone-like spreading of
  correlations in a quantum many-body system},}\ }}\href {\doibase
  10.1038/nature10748} {\bibfield  {journal} {\bibinfo  {journal} {Nature}\
  }\textbf {\bibinfo {volume} {481}},\ \bibinfo {pages} {484--487} (\bibinfo
  {year} {2012})}\BibitemShut {NoStop}%
\bibitem [{\citenamefont {Trotzky}\ \emph {et~al.}(2012)\citenamefont
  {Trotzky}, \citenamefont {Chen}, \citenamefont {Flesch}, \citenamefont
  {{McCulloch}}, \citenamefont {Schollw\"ock}, \citenamefont {Eisert},\ and\
  \citenamefont {Bloch}}]{trotzky_probing_2012}%
  \BibitemOpen
  \bibfield  {author} {\bibinfo {author} {\bibfnamefont {S.}~\bibnamefont
  {Trotzky}}, \bibinfo {author} {\bibfnamefont {Y.-A.}\ \bibnamefont {Chen}},
  \bibinfo {author} {\bibfnamefont {A.}~\bibnamefont {Flesch}}, \bibinfo
  {author} {\bibfnamefont {I.~P.}\ \bibnamefont {{McCulloch}}}, \bibinfo
  {author} {\bibfnamefont {U.}~\bibnamefont {Schollw\"ock}}, \bibinfo {author}
  {\bibfnamefont {J.}~\bibnamefont {Eisert}}, \ and\ \bibinfo {author}
  {\bibfnamefont {I.}~\bibnamefont {Bloch}},\ }\bibfield  {title}
  {{\selectlanguage {english}\enquote {\bibinfo {title} {Probing the relaxation
  towards equilibrium in an isolated strongly correlated one-dimensional {Bose}
  gas},}\ }}\href {\doibase 10.1038/nphys2232} {\bibfield  {journal} {\bibinfo
  {journal} {Nat. Phys.}\ ,\ \bibinfo {pages} {325--330}} (\bibinfo {year}
  {2012})}\BibitemShut {NoStop}%
\bibitem [{\citenamefont {Schuch}\ \emph {et~al.}(2008)\citenamefont {Schuch},
  \citenamefont {Wolf}, \citenamefont {Verstraete},\ and\ \citenamefont
  {Cirac}}]{schuch_entropy_2008}%
  \BibitemOpen
  \bibfield  {author} {\bibinfo {author} {\bibfnamefont {N.}~\bibnamefont
  {Schuch}}, \bibinfo {author} {\bibfnamefont {M.~M.}\ \bibnamefont {Wolf}},
  \bibinfo {author} {\bibfnamefont {F.}~\bibnamefont {Verstraete}}, \ and\
  \bibinfo {author} {\bibfnamefont {J.~I.}\ \bibnamefont {Cirac}},\ }\bibfield
  {title} {\enquote {\bibinfo {title} {Entropy scaling and simulability by
  matrix product states},}\ }\href {\doibase 10.1103/PhysRevLett.100.030504}
  {\bibfield  {journal} {\bibinfo  {journal} {Phys. Rev. Lett.}\ }\textbf
  {\bibinfo {volume} {100}},\ \bibinfo {pages} {030504} (\bibinfo {year}
  {2008})}\BibitemShut {NoStop}%
\bibitem [{\citenamefont {Verstraete}\ and\ \citenamefont
  {Cirac}(2006)}]{verstraete_matrix_2006}%
  \BibitemOpen
  \bibfield  {author} {\bibinfo {author} {\bibfnamefont {F.}~\bibnamefont
  {Verstraete}}\ and\ \bibinfo {author} {\bibfnamefont {J.~I.}\ \bibnamefont
  {Cirac}},\ }\bibfield  {title} {\enquote {\bibinfo {title} {Matrix product
  states represent ground states faithfully},}\ }\href {\doibase
  10.1103/PhysRevB.73.094423} {\bibfield  {journal} {\bibinfo  {journal} {Phys.
  Rev. B}\ }\textbf {\bibinfo {volume} {73}},\ \bibinfo {pages} {094423}
  (\bibinfo {year} {2006})}\BibitemShut {NoStop}%
\bibitem [{\citenamefont {Cirac}\ and\ \citenamefont
  {Zoller}(2012)}]{cirac_goals_2012}%
  \BibitemOpen
  \bibfield  {author} {\bibinfo {author} {\bibfnamefont {J.~I.}\ \bibnamefont
  {Cirac}}\ and\ \bibinfo {author} {\bibfnamefont {P.}~\bibnamefont {Zoller}},\
  }\bibfield  {title} {{\selectlanguage {english}\enquote {\bibinfo {title}
  {Goals and opportunities in quantum simulation},}\ }}\href {\doibase
  10.1038/nphys2275} {\bibfield  {journal} {\bibinfo  {journal} {Nat. Phys.}\
  }\textbf {\bibinfo {volume} {8}},\ \bibinfo {pages} {264--266} (\bibinfo
  {year} {2012})}\BibitemShut {NoStop}%
\bibitem [{\citenamefont {H\"affner}\ \emph {et~al.}(2008)\citenamefont
  {H\"affner}, \citenamefont {Roos},\ and\ \citenamefont
  {Blatt}}]{haffner_quantum_2008}%
  \BibitemOpen
  \bibfield  {author} {\bibinfo {author} {\bibfnamefont {H.}~\bibnamefont
  {H\"affner}}, \bibinfo {author} {\bibfnamefont {C.}~\bibnamefont {Roos}}, \
  and\ \bibinfo {author} {\bibfnamefont {R.}~\bibnamefont {Blatt}},\ }\bibfield
   {title} {\enquote {\bibinfo {title} {Quantum computing with trapped ions},}\
  }\href {\doibase 10.1016/j.physrep.2008.09.003} {\bibfield  {journal}
  {\bibinfo  {journal} {Phys. Rep.}\ }\textbf {\bibinfo {volume} {469}},\
  \bibinfo {pages} {155--203} (\bibinfo {year} {2008})}\BibitemShut {NoStop}%
\bibitem [{\citenamefont {Wineland}\ and\ \citenamefont
  {Leibfried}(2011)}]{wineland_quantum_2011}%
  \BibitemOpen
  \bibfield  {author} {\bibinfo {author} {\bibfnamefont {D.~J.}\ \bibnamefont
  {Wineland}}\ and\ \bibinfo {author} {\bibfnamefont {D.}~\bibnamefont
  {Leibfried}},\ }\bibfield  {title} {{\selectlanguage {english}\enquote
  {\bibinfo {title} {Quantum information processing and metrology with trapped
  ions},}\ }}\href {\doibase 10.1002/lapl.201010125} {\bibfield  {journal}
  {\bibinfo  {journal} {Laser Phys. Lett.}\ }\textbf {\bibinfo {volume} {8}},\
  \bibinfo {pages} {175Ð188} (\bibinfo {year} {2011})}\BibitemShut {NoStop}%
\bibitem [{\citenamefont {Lanyon}\ \emph {et~al.}(2011)\citenamefont {Lanyon},
  \citenamefont {Hempel}, \citenamefont {Nigg}, \citenamefont {M\"uller},
  \citenamefont {Gerritsma}, \citenamefont {Z\"ahringer}, \citenamefont
  {Schindler}, \citenamefont {Barreiro}, \citenamefont {Rambach}, \citenamefont
  {Kirchmair}, \citenamefont {Hennrich}, \citenamefont {Zoller}, \citenamefont
  {Blatt},\ and\ \citenamefont {Roos}}]{lanyon_universal_2011}%
  \BibitemOpen
  \bibfield  {author} {\bibinfo {author} {\bibfnamefont {B.~P.}\ \bibnamefont
  {Lanyon}}, \bibinfo {author} {\bibfnamefont {C.}~\bibnamefont {Hempel}},
  \bibinfo {author} {\bibfnamefont {D.}~\bibnamefont {Nigg}}, \bibinfo {author}
  {\bibfnamefont {M.}~\bibnamefont {M\"uller}}, \bibinfo {author}
  {\bibfnamefont {R.}~\bibnamefont {Gerritsma}}, \bibinfo {author}
  {\bibfnamefont {F.}~\bibnamefont {Z\"ahringer}}, \bibinfo {author}
  {\bibfnamefont {P.}~\bibnamefont {Schindler}}, \bibinfo {author}
  {\bibfnamefont {J.~T.}\ \bibnamefont {Barreiro}}, \bibinfo {author}
  {\bibfnamefont {M.}~\bibnamefont {Rambach}}, \bibinfo {author} {\bibfnamefont
  {G.}~\bibnamefont {Kirchmair}}, \bibinfo {author} {\bibfnamefont
  {M.}~\bibnamefont {Hennrich}}, \bibinfo {author} {\bibfnamefont
  {P.}~\bibnamefont {Zoller}}, \bibinfo {author} {\bibfnamefont
  {R.}~\bibnamefont {Blatt}}, \ and\ \bibinfo {author} {\bibfnamefont {C.~F.}\
  \bibnamefont {Roos}},\ }\bibfield  {title} {{\selectlanguage
  {english}\enquote {\bibinfo {title} {Universal digital quantum simulation
  with trapped ions},}\ }}\href {\doibase 10.1126/science.1208001} {\bibfield
  {journal} {\bibinfo  {journal} {Science}\ }\textbf {\bibinfo {volume}
  {334}},\ \bibinfo {pages} {57--61} (\bibinfo {year} {2011})}\BibitemShut
  {NoStop}%
\bibitem [{\citenamefont {Roos}\ \emph {et~al.}(2004)\citenamefont {Roos},
  \citenamefont {Lancaster}, \citenamefont {Riebe}, \citenamefont
  {H{\"a}ffner}, \citenamefont {H{\"a}nsel}, \citenamefont {Gulde},
  \citenamefont {Becher}, \citenamefont {Eschner}, \citenamefont
  {Schmidt-Kaler},\ and\ \citenamefont {Blatt}}]{roos_bell_2004}%
  \BibitemOpen
  \bibfield  {author} {\bibinfo {author} {\bibfnamefont {C.~F.}\ \bibnamefont
  {Roos}}, \bibinfo {author} {\bibfnamefont {G.~P.~T.}\ \bibnamefont
  {Lancaster}}, \bibinfo {author} {\bibfnamefont {M.}~\bibnamefont {Riebe}},
  \bibinfo {author} {\bibfnamefont {H.}~\bibnamefont {H{\"a}ffner}}, \bibinfo
  {author} {\bibfnamefont {W.}~\bibnamefont {H{\"a}nsel}}, \bibinfo {author}
  {\bibfnamefont {S.}~\bibnamefont {Gulde}}, \bibinfo {author} {\bibfnamefont
  {C.}~\bibnamefont {Becher}}, \bibinfo {author} {\bibfnamefont
  {J.}~\bibnamefont {Eschner}}, \bibinfo {author} {\bibfnamefont
  {F.}~\bibnamefont {Schmidt-Kaler}}, \ and\ \bibinfo {author} {\bibfnamefont
  {R.}~\bibnamefont {Blatt}},\ }\bibfield  {title} {\enquote {\bibinfo {title}
  {Bell states of atoms with ultralong lifetimes and their tomographic state
  analysis},}\ }\href {\doibase 10.1103/PhysRevLett.92.220402} {\bibfield
  {journal} {\bibinfo  {journal} {Phys. Rev. Lett.}\ }\textbf {\bibinfo
  {volume} {92}},\ \bibinfo {pages} {220402} (\bibinfo {year}
  {2004})}\BibitemShut {NoStop}%
\bibitem [{\citenamefont {Porras}\ and\ \citenamefont
  {Cirac}(2004)}]{porras_effective_2004}%
  \BibitemOpen
  \bibfield  {author} {\bibinfo {author} {\bibfnamefont {D.}~\bibnamefont
  {Porras}}\ and\ \bibinfo {author} {\bibfnamefont {J.~I.}\ \bibnamefont
  {Cirac}},\ }\bibfield  {title} {\enquote {\bibinfo {title} {Effective quantum
  spin systems with trapped ions},}\ }\href {\doibase
  10.1103/PhysRevLett.92.207901} {\bibfield  {journal} {\bibinfo  {journal}
  {Phys. Rev. Lett.}\ }\textbf {\bibinfo {volume} {92}},\ \bibinfo {pages}
  {207901} (\bibinfo {year} {2004})}\BibitemShut {NoStop}%
\bibitem [{\citenamefont {Schneider}\ \emph {et~al.}(2012)\citenamefont
  {Schneider}, \citenamefont {Porras},\ and\ \citenamefont
  {Schaetz}}]{schneider_experimental_2012}%
  \BibitemOpen
  \bibfield  {author} {\bibinfo {author} {\bibfnamefont {C.}~\bibnamefont
  {Schneider}}, \bibinfo {author} {\bibfnamefont {D.}~\bibnamefont {Porras}}, \
  and\ \bibinfo {author} {\bibfnamefont {T.}~\bibnamefont {Schaetz}},\
  }\bibfield  {title} {{\selectlanguage {english}\enquote {\bibinfo {title}
  {Experimental quantum simulations of many-body physics with trapped ions},}\
  }}\href {\doibase 10.1088/0034-4885/75/2/024401} {\bibfield  {journal}
  {\bibinfo  {journal} {Rep. Prog. Phys}\ }\textbf {\bibinfo {volume} {75}},\
  \bibinfo {pages} {024401} (\bibinfo {year} {2012})}\BibitemShut {NoStop}%
\bibitem [{\citenamefont {Friedenauer}\ \emph {et~al.}(2008)\citenamefont
  {Friedenauer}, \citenamefont {Schmitz}, \citenamefont {Glueckert},
  \citenamefont {Porras},\ and\ \citenamefont
  {Schaetz}}]{friedenauer_simulating_2008}%
  \BibitemOpen
  \bibfield  {author} {\bibinfo {author} {\bibfnamefont {A.}~\bibnamefont
  {Friedenauer}}, \bibinfo {author} {\bibfnamefont {H.}~\bibnamefont
  {Schmitz}}, \bibinfo {author} {\bibfnamefont {J.~T.}\ \bibnamefont
  {Glueckert}}, \bibinfo {author} {\bibfnamefont {D.}~\bibnamefont {Porras}}, \
  and\ \bibinfo {author} {\bibfnamefont {T.}~\bibnamefont {Schaetz}},\
  }\bibfield  {title} {{\selectlanguage {english}\enquote {\bibinfo {title}
  {Simulating a quantum magnet with trapped ions},}\ }}\href {\doibase
  10.1038/nphys1032} {\bibfield  {journal} {\bibinfo  {journal} {Nat. Phys.}\
  }\textbf {\bibinfo {volume} {4}},\ \bibinfo {pages} {757--761} (\bibinfo
  {year} {2008})}\BibitemShut {NoStop}%
\bibitem [{\citenamefont {Islam}\ \emph {et~al.}(2011)\citenamefont {Islam},
  \citenamefont {Edwards}, \citenamefont {Kim}, \citenamefont {Korenblit},
  \citenamefont {Noh}, \citenamefont {Carmichael}, \citenamefont {Lin},
  \citenamefont {Duan}, \citenamefont {Wang}, \citenamefont {Freericks},\ and\
  \citenamefont {Monroe}}]{islam_onset_2011}%
  \BibitemOpen
  \bibfield  {author} {\bibinfo {author} {\bibfnamefont {R.}~\bibnamefont
  {Islam}}, \bibinfo {author} {\bibfnamefont {E.~E.}\ \bibnamefont {Edwards}},
  \bibinfo {author} {\bibfnamefont {K.}~\bibnamefont {Kim}}, \bibinfo {author}
  {\bibfnamefont {S.}~\bibnamefont {Korenblit}}, \bibinfo {author}
  {\bibfnamefont {C.}~\bibnamefont {Noh}}, \bibinfo {author} {\bibfnamefont
  {H.}~\bibnamefont {Carmichael}}, \bibinfo {author} {\bibfnamefont {G.-D.}\
  \bibnamefont {Lin}}, \bibinfo {author} {\bibfnamefont {L.-M.}\ \bibnamefont
  {Duan}}, \bibinfo {author} {\bibfnamefont {C.-C.~J.}\ \bibnamefont {Wang}},
  \bibinfo {author} {\bibfnamefont {J.~K.}\ \bibnamefont {Freericks}}, \ and\
  \bibinfo {author} {\bibfnamefont {C.}~\bibnamefont {Monroe}},\ }\bibfield
  {title} {{\selectlanguage {english}\enquote {\bibinfo {title} {Onset of a
  quantum phase transition with a trapped ion quantum simulator},}\ }}\href
  {\doibase 10.1038/ncomms1374} {\bibfield  {journal} {\bibinfo  {journal}
  {Nat. Comm.}\ }\textbf {\bibinfo {volume} {2}},\ \bibinfo {pages} {377}
  (\bibinfo {year} {2011})}\BibitemShut {NoStop}%
\bibitem [{\citenamefont {Kim}\ \emph {et~al.}(2011)\citenamefont {Kim},
  \citenamefont {Korenblit}, \citenamefont {Islam}, \citenamefont {Edwards},
  \citenamefont {Chang}, \citenamefont {Noh}, \citenamefont {Carmichael},
  \citenamefont {Lin}, \citenamefont {Duan}, \citenamefont {Wang},
  \citenamefont {Freericks},\ and\ \citenamefont {Monroe}}]{kim_quantum_2011}%
  \BibitemOpen
  \bibfield  {author} {\bibinfo {author} {\bibfnamefont {K.}~\bibnamefont
  {Kim}}, \bibinfo {author} {\bibfnamefont {S.}~\bibnamefont {Korenblit}},
  \bibinfo {author} {\bibfnamefont {R.}~\bibnamefont {Islam}}, \bibinfo
  {author} {\bibfnamefont {E.~E.}\ \bibnamefont {Edwards}}, \bibinfo {author}
  {\bibfnamefont {M.-S.}\ \bibnamefont {Chang}}, \bibinfo {author}
  {\bibfnamefont {C.}~\bibnamefont {Noh}}, \bibinfo {author} {\bibfnamefont
  {H.}~\bibnamefont {Carmichael}}, \bibinfo {author} {\bibfnamefont {G.-D.}\
  \bibnamefont {Lin}}, \bibinfo {author} {\bibfnamefont {L.-M.}\ \bibnamefont
  {Duan}}, \bibinfo {author} {\bibfnamefont {C.-C.~J.}\ \bibnamefont {Wang}},
  \bibinfo {author} {\bibfnamefont {J.~K.}\ \bibnamefont {Freericks}}, \ and\
  \bibinfo {author} {\bibfnamefont {C.}~\bibnamefont {Monroe}},\ }\bibfield
  {title} {{\selectlanguage {english}\enquote {\bibinfo {title} {Quantum
  simulation of the transverse {Ising} model with trapped ions},}\ }}\href
  {\doibase 10.1088/1367-2630/13/10/105003} {\bibfield  {journal} {\bibinfo
  {journal} {New J. Phys.}\ }\textbf {\bibinfo {volume} {13}},\ \bibinfo
  {pages} {105003} (\bibinfo {year} {2011})}\BibitemShut {NoStop}%
\bibitem [{\citenamefont {Britton}\ \emph {et~al.}(2012)\citenamefont
  {Britton}, \citenamefont {Sawyer}, \citenamefont {Keith}, \citenamefont
  {Wang}, \citenamefont {Freericks}, \citenamefont {Uys}, \citenamefont
  {Biercuk},\ and\ \citenamefont {John J.~Bollinger}}]{Britton:2012}%
  \BibitemOpen
  \bibfield  {author} {\bibinfo {author} {\bibfnamefont {J.~W.}\ \bibnamefont
  {Britton}}, \bibinfo {author} {\bibfnamefont {B.~C.}\ \bibnamefont {Sawyer}},
  \bibinfo {author} {\bibfnamefont {A.~C.}\ \bibnamefont {Keith}}, \bibinfo
  {author} {\bibfnamefont {C.-C.~J.}\ \bibnamefont {Wang}}, \bibinfo {author}
  {\bibfnamefont {J.~K.}\ \bibnamefont {Freericks}}, \bibinfo {author}
  {\bibfnamefont {H.}~\bibnamefont {Uys}}, \bibinfo {author} {\bibfnamefont
  {M.~J.}\ \bibnamefont {Biercuk}}, \ and\ \bibinfo {author} {\bibfnamefont
  {J.~J.}\ \bibnamefont {John J.~Bollinger}},\ }\bibfield  {title} {\enquote
  {\bibinfo {title} {Engineered two-dimensional {Ising} interactions in a
  trapped-ion quantum simulator with hundreds of spins},}\ }\href@noop {}
  {\bibfield  {journal} {\bibinfo  {journal} {Nature}\ }\textbf {\bibinfo
  {volume} {484}},\ \bibinfo {pages} {489--492} (\bibinfo {year}
  {2012})}\BibitemShut {NoStop}%
\bibitem [{\citenamefont {Islam}\ \emph {et~al.}(2013)\citenamefont {Islam},
  \citenamefont {Senko}, \citenamefont {Campbell}, \citenamefont {Korenblit},
  \citenamefont {Smith}, \citenamefont {Lee}, \citenamefont {Edwards},
  \citenamefont {Wang}, \citenamefont {Freericks},\ and\ \citenamefont
  {Monroe}}]{islam_emergence_2013}%
  \BibitemOpen
  \bibfield  {author} {\bibinfo {author} {\bibfnamefont {R.}~\bibnamefont
  {Islam}}, \bibinfo {author} {\bibfnamefont {C.}~\bibnamefont {Senko}},
  \bibinfo {author} {\bibfnamefont {W.~C.}\ \bibnamefont {Campbell}}, \bibinfo
  {author} {\bibfnamefont {S.}~\bibnamefont {Korenblit}}, \bibinfo {author}
  {\bibfnamefont {J.}~\bibnamefont {Smith}}, \bibinfo {author} {\bibfnamefont
  {A.}~\bibnamefont {Lee}}, \bibinfo {author} {\bibfnamefont {E.~E.}\
  \bibnamefont {Edwards}}, \bibinfo {author} {\bibfnamefont {C.-C.~J.}\
  \bibnamefont {Wang}}, \bibinfo {author} {\bibfnamefont {J.~K.}\ \bibnamefont
  {Freericks}}, \ and\ \bibinfo {author} {\bibfnamefont {C.}~\bibnamefont
  {Monroe}},\ }\bibfield  {title} {\enquote {\bibinfo {title} {Emergence and
  frustration of magnetic order with variable-range interactions in a trapped
  ion quantum simulator},}\ }\href@noop {} {\bibfield  {journal} {\bibinfo
  {journal} {Science}\ }\textbf {\bibinfo {volume} {340}},\ \bibinfo {pages}
  {583--587} (\bibinfo {year} {2013})}\BibitemShut {NoStop}%
\bibitem [{\citenamefont {Richerme}\ \emph {et~al.}(2013)\citenamefont
  {Richerme}, \citenamefont {Senko}, \citenamefont {Korenblit}, \citenamefont
  {Smith}, \citenamefont {Lee}, \citenamefont {Islam}, \citenamefont
  {Campbell},\ and\ \citenamefont {Monroe}}]{richerme_trapped-ion_2013}%
  \BibitemOpen
  \bibfield  {author} {\bibinfo {author} {\bibfnamefont {P.}~\bibnamefont
  {Richerme}}, \bibinfo {author} {\bibfnamefont {C.}~\bibnamefont {Senko}},
  \bibinfo {author} {\bibfnamefont {S.}~\bibnamefont {Korenblit}}, \bibinfo
  {author} {\bibfnamefont {J.}~\bibnamefont {Smith}}, \bibinfo {author}
  {\bibfnamefont {A.}~\bibnamefont {Lee}}, \bibinfo {author} {\bibfnamefont
  {R.}~\bibnamefont {Islam}}, \bibinfo {author} {\bibfnamefont {W.~C.}\
  \bibnamefont {Campbell}}, \ and\ \bibinfo {author} {\bibfnamefont
  {C.}~\bibnamefont {Monroe}},\ }\bibfield  {title} {\enquote {\bibinfo {title}
  {Quantum catalysis of magnetic phase transitions in a quantum simulator},}\
  }\href {http://arxiv.org/abs/1303.6983} {\bibfield  {journal} {\bibinfo
  {journal} {arXiv:1303.6983}\ } (\bibinfo {year} {2013})}\BibitemShut
  {NoStop}%
\bibitem [{\citenamefont {Kim}\ \emph {et~al.}(2009)\citenamefont {Kim},
  \citenamefont {Chang}, \citenamefont {Islam}, \citenamefont {Korenblit},
  \citenamefont {Duan},\ and\ \citenamefont {Monroe}}]{PhysRevLett.103.120502}%
  \BibitemOpen
  \bibfield  {author} {\bibinfo {author} {\bibfnamefont {K.}~\bibnamefont
  {Kim}}, \bibinfo {author} {\bibfnamefont {M.-S.}\ \bibnamefont {Chang}},
  \bibinfo {author} {\bibfnamefont {R.}~\bibnamefont {Islam}}, \bibinfo
  {author} {\bibfnamefont {S.}~\bibnamefont {Korenblit}}, \bibinfo {author}
  {\bibfnamefont {L.-M.}\ \bibnamefont {Duan}}, \ and\ \bibinfo {author}
  {\bibfnamefont {C.}~\bibnamefont {Monroe}},\ }\bibfield  {title} {\enquote
  {\bibinfo {title} {Entanglement and tunable spin-spin couplings between
  trapped ions using multiple transverse modes},}\ }\href {\doibase
  10.1103/PhysRevLett.103.120502} {\bibfield  {journal} {\bibinfo  {journal}
  {Phys. Rev. Lett.}\ }\textbf {\bibinfo {volume} {103}},\ \bibinfo {pages}
  {120502} (\bibinfo {year} {2009})}\BibitemShut {NoStop}%
\bibitem [{\citenamefont {Korenblit}\ \emph {et~al.}(2012)\citenamefont
  {Korenblit}, \citenamefont {Kafri}, \citenamefont {Campbell}, \citenamefont
  {Islam}, \citenamefont {Edwards}, \citenamefont {Gong}, \citenamefont {Lin},
  \citenamefont {Duan}, \citenamefont {Kim}, \citenamefont {Kim},\ and\
  \citenamefont {Monroe}}]{1367-2630-14-9-095024}%
  \BibitemOpen
  \bibfield  {author} {\bibinfo {author} {\bibfnamefont {S.}~\bibnamefont
  {Korenblit}}, \bibinfo {author} {\bibfnamefont {D.}~\bibnamefont {Kafri}},
  \bibinfo {author} {\bibfnamefont {W.~C.}\ \bibnamefont {Campbell}}, \bibinfo
  {author} {\bibfnamefont {R.}~\bibnamefont {Islam}}, \bibinfo {author}
  {\bibfnamefont {E.~E.}\ \bibnamefont {Edwards}}, \bibinfo {author}
  {\bibfnamefont {Z.-X.}\ \bibnamefont {Gong}}, \bibinfo {author}
  {\bibfnamefont {G.-D.}\ \bibnamefont {Lin}}, \bibinfo {author} {\bibfnamefont
  {L.-M.}\ \bibnamefont {Duan}}, \bibinfo {author} {\bibfnamefont
  {J.}~\bibnamefont {Kim}}, \bibinfo {author} {\bibfnamefont {K.}~\bibnamefont
  {Kim}}, \ and\ \bibinfo {author} {\bibfnamefont {C.}~\bibnamefont {Monroe}},\
  }\bibfield  {title} {\enquote {\bibinfo {title} {Quantum simulation of spin
  models on an arbitrary lattice with trapped ions},}\ }\href@noop {}
  {\bibfield  {journal} {\bibinfo  {journal} {New J. Phys.}\ }\textbf {\bibinfo
  {volume} {14}},\ \bibinfo {pages} {095024} (\bibinfo {year}
  {2012})}\BibitemShut {NoStop}%
\bibitem [{\citenamefont {Wang}\ \emph {et~al.}(2013)\citenamefont {Wang},
  \citenamefont {Keith},\ and\ \citenamefont {Freericks}}]{PhysRevA.87.013422}%
  \BibitemOpen
  \bibfield  {author} {\bibinfo {author} {\bibfnamefont {C.-C.~J.}\
  \bibnamefont {Wang}}, \bibinfo {author} {\bibfnamefont {A.~C.}\ \bibnamefont
  {Keith}}, \ and\ \bibinfo {author} {\bibfnamefont {J.~K.}\ \bibnamefont
  {Freericks}},\ }\bibfield  {title} {\enquote {\bibinfo {title}
  {Phonon-mediated quantum spin simulator employing a planar ionic crystal in a
  {Penning} trap},}\ }\href {\doibase 10.1103/PhysRevA.87.013422} {\bibfield
  {journal} {\bibinfo  {journal} {Phys. Rev. A}\ }\textbf {\bibinfo {volume}
  {87}},\ \bibinfo {pages} {013422} (\bibinfo {year} {2013})}\BibitemShut
  {NoStop}%
\bibitem [{\citenamefont {Crosswhite}\ \emph {et~al.}(2008)\citenamefont
  {Crosswhite}, \citenamefont {Doherty},\ and\ \citenamefont
  {Vidal}}]{crosswhite_applying_2008}%
  \BibitemOpen
  \bibfield  {author} {\bibinfo {author} {\bibfnamefont {G.~M.}\ \bibnamefont
  {Crosswhite}}, \bibinfo {author} {\bibfnamefont {A.~C.}\ \bibnamefont
  {Doherty}}, \ and\ \bibinfo {author} {\bibfnamefont {G.}~\bibnamefont
  {Vidal}},\ }\bibfield  {title} {\enquote {\bibinfo {title} {Applying matrix
  product operators to model systems with long-range interactions},}\ }\href
  {\doibase 10.1103/PhysRevB.78.035116} {\bibfield  {journal} {\bibinfo
  {journal} {Phys. Rev. B}\ }\textbf {\bibinfo {volume} {78}},\ \bibinfo
  {pages} {035116} (\bibinfo {year} {2008})}\BibitemShut {NoStop}%
\bibitem [{\citenamefont {Pirvu}\ \emph {et~al.}(2010)\citenamefont {Pirvu},
  \citenamefont {Murg}, \citenamefont {Cirac},\ and\ \citenamefont
  {Verstraete}}]{pirvu_matrix_2010}%
  \BibitemOpen
  \bibfield  {author} {\bibinfo {author} {\bibfnamefont {B.}~\bibnamefont
  {Pirvu}}, \bibinfo {author} {\bibfnamefont {V.}~\bibnamefont {Murg}},
  \bibinfo {author} {\bibfnamefont {J.~I.}\ \bibnamefont {Cirac}}, \ and\
  \bibinfo {author} {\bibfnamefont {F.}~\bibnamefont {Verstraete}},\ }\bibfield
   {title} {\enquote {\bibinfo {title} {Matrix product operator
  representations},}\ }\href {\doibase 10.1088/1367-2630/12/2/025012}
  {\bibfield  {journal} {\bibinfo  {journal} {New J. Phys.}\ }\textbf {\bibinfo
  {volume} {12}},\ \bibinfo {pages} {025012} (\bibinfo {year}
  {2010})}\BibitemShut {NoStop}%
\bibitem [{\citenamefont {Fr\"owis}\ \emph {et~al.}(2010)\citenamefont
  {Fr\"owis}, \citenamefont {Nebendahl},\ and\ \citenamefont
  {D\"ur}}]{frowis_tensor_2010}%
  \BibitemOpen
  \bibfield  {author} {\bibinfo {author} {\bibfnamefont {F.}~\bibnamefont
  {Fr\"owis}}, \bibinfo {author} {\bibfnamefont {V.}~\bibnamefont {Nebendahl}},
  \ and\ \bibinfo {author} {\bibfnamefont {W.}~\bibnamefont {D\"ur}},\
  }\bibfield  {title} {\enquote {\bibinfo {title} {Tensor operators:
  Constructions and applications for long-range interaction systems},}\ }\href
  {\doibase 10.1103/PhysRevA.81.062337} {\bibfield  {journal} {\bibinfo
  {journal} {Phys. Rev. A}\ }\textbf {\bibinfo {volume} {81}},\ \bibinfo
  {pages} {062337} (\bibinfo {year} {2010})}\BibitemShut {NoStop}%
\bibitem [{\citenamefont {Vidal}(2004)}]{vidal_efficient_2004}%
  \BibitemOpen
  \bibfield  {author} {\bibinfo {author} {\bibfnamefont {G.}~\bibnamefont
  {Vidal}},\ }\bibfield  {title} {\enquote {\bibinfo {title} {Efficient
  simulation of one-dimensional quantum many-body systems},}\ }\href {\doibase
  10.1103/PhysRevLett.93.040502} {\bibfield  {journal} {\bibinfo  {journal}
  {Phys. Rev. Lett.}\ }\textbf {\bibinfo {volume} {93}},\ \bibinfo {pages}
  {040502} (\bibinfo {year} {2004})}\BibitemShut {NoStop}%
\bibitem [{\citenamefont {White}\ and\ \citenamefont
  {Feiguin}(2004)}]{white_real-time_2004}%
  \BibitemOpen
  \bibfield  {author} {\bibinfo {author} {\bibfnamefont {S.~R.}\ \bibnamefont
  {White}}\ and\ \bibinfo {author} {\bibfnamefont {A.~E.}\ \bibnamefont
  {Feiguin}},\ }\bibfield  {title} {\enquote {\bibinfo {title} {Real-time
  evolution using the density matrix renormalization group},}\ }\href {\doibase
  10.1103/PhysRevLett.93.076401} {\bibfield  {journal} {\bibinfo  {journal}
  {Phys. Rev. Lett.}\ }\textbf {\bibinfo {volume} {93}},\ \bibinfo {pages}
  {076401} (\bibinfo {year} {2004})}\BibitemShut {NoStop}%
\bibitem [{\citenamefont {Daley}\ \emph {et~al.}(2004)\citenamefont {Daley},
  \citenamefont {Kollath}, \citenamefont {Schollw{\"o}ck},\ and\ \citenamefont
  {Vidal}}]{daley_time-dependent_2004}%
  \BibitemOpen
  \bibfield  {author} {\bibinfo {author} {\bibfnamefont {A.~J.}\ \bibnamefont
  {Daley}}, \bibinfo {author} {\bibfnamefont {C.}~\bibnamefont {Kollath}},
  \bibinfo {author} {\bibfnamefont {U.}~\bibnamefont {Schollw{\"o}ck}}, \ and\
  \bibinfo {author} {\bibfnamefont {G.}~\bibnamefont {Vidal}},\ }\bibfield
  {title} {\enquote {\bibinfo {title} {Time-dependent density-matrix
  renormalization-group using adaptive effective {Hilbert} spaces},}\ }\href
  {\doibase 10.1088/1742-5468/2004/04/P04005} {\bibfield  {journal} {\bibinfo
  {journal} {J. Stat. Mech. Theor. Exp.}\ }\textbf {\bibinfo {volume} {2004}},\
  \bibinfo {pages} {P04005} (\bibinfo {year} {2004})}\BibitemShut {NoStop}%
\bibitem [{\citenamefont
  {Schollw\"ock}(2011)}]{schollwock_density-matrix_2011}%
  \BibitemOpen
  \bibfield  {author} {\bibinfo {author} {\bibfnamefont {U.}~\bibnamefont
  {Schollw\"ock}},\ }\bibfield  {title} {\enquote {\bibinfo {title} {The
  density-matrix renormalization group in the age of matrix product states},}\
  }\href {\doibase 10.1016/j.aop.2010.09.012} {\bibfield  {journal} {\bibinfo
  {journal} {Ann. Phys.}\ }\textbf {\bibinfo {volume} {326}},\ \bibinfo {pages}
  {96--192} (\bibinfo {year} {2011})}\BibitemShut {NoStop}%
\bibitem [{\citenamefont {Verstraete}\ \emph {et~al.}(2008)\citenamefont
  {Verstraete}, \citenamefont {Murg},\ and\ \citenamefont
  {Cirac}}]{verstraete_matrix_2008}%
  \BibitemOpen
  \bibfield  {author} {\bibinfo {author} {\bibfnamefont {F.}~\bibnamefont
  {Verstraete}}, \bibinfo {author} {\bibfnamefont {V.}~\bibnamefont {Murg}}, \
  and\ \bibinfo {author} {\bibfnamefont {J.~I.}\ \bibnamefont {Cirac}},\
  }\bibfield  {title} {\enquote {\bibinfo {title} {Matrix product states,
  projected entangled pair states, and variational renormalization group
  methods for quantum spin systems},}\ }\href {\doibase
  10.1080/14789940801912366} {\bibfield  {journal} {\bibinfo  {journal} {Adv.
  Phys.}\ }\textbf {\bibinfo {volume} {57}},\ \bibinfo {pages} {143--224}
  (\bibinfo {year} {2008})}\BibitemShut {NoStop}%
\bibitem [{\citenamefont {Lieb}\ and\ \citenamefont
  {Robinson}(1972)}]{lieb_finite_1972}%
  \BibitemOpen
  \bibfield  {author} {\bibinfo {author} {\bibfnamefont {E.~H.}\ \bibnamefont
  {Lieb}}\ and\ \bibinfo {author} {\bibfnamefont {D.~W.}\ \bibnamefont
  {Robinson}},\ }\bibfield  {title} {{\selectlanguage {english}\enquote
  {\bibinfo {title} {The finite group velocity of quantum spin systems},}\
  }}\href {\doibase 10.1007/BF01645779} {\bibfield  {journal} {\bibinfo
  {journal} {Communications in Mathematical Physics}\ }\textbf {\bibinfo
  {volume} {28}},\ \bibinfo {pages} {251--257} (\bibinfo {year}
  {1972})}\BibitemShut {NoStop}%
\bibitem [{\citenamefont {Calabrese}\ and\ \citenamefont
  {Cardy}(2005)}]{calabrese_evolution_2005}%
  \BibitemOpen
  \bibfield  {author} {\bibinfo {author} {\bibfnamefont {P.}~\bibnamefont
  {Calabrese}}\ and\ \bibinfo {author} {\bibfnamefont {J.}~\bibnamefont
  {Cardy}},\ }\bibfield  {title} {\enquote {\bibinfo {title} {Evolution of
  entanglement entropy in one-dimensional systems},}\ }\href {\doibase
  10.1088/1742-5468/2005/04/P04010} {\bibfield  {journal} {\bibinfo  {journal}
  {J. Stat. Mech. Theor. Exp.}\ }\textbf {\bibinfo {volume} {2005}},\ \bibinfo
  {pages} {P04010} (\bibinfo {year} {2005})}\BibitemShut {NoStop}%
\bibitem [{\citenamefont {Cramer}\ \emph {et~al.}(2008)\citenamefont {Cramer},
  \citenamefont {Serafini},\ and\ \citenamefont
  {Eisert}}]{cramer_locality_2008}%
  \BibitemOpen
  \bibfield  {author} {\bibinfo {author} {\bibfnamefont {M.}~\bibnamefont
  {Cramer}}, \bibinfo {author} {\bibfnamefont {A.}~\bibnamefont {Serafini}}, \
  and\ \bibinfo {author} {\bibfnamefont {J.}~\bibnamefont {Eisert}},\
  }\bibfield  {title} {\enquote {\bibinfo {title} {Locality of dynamics in
  general harmonic quantum systems},}\ }\href@noop {} {\bibfield  {journal}
  {\bibinfo  {journal} {{arXiv:0803.0890}}\ } (\bibinfo {year}
  {2008})}\BibitemShut {NoStop}%
\bibitem [{\citenamefont {Vidal}\ \emph {et~al.}(2004)\citenamefont {Vidal},
  \citenamefont {Palacios},\ and\ \citenamefont
  {Aslangul}}]{vidal_entanglement_2004}%
  \BibitemOpen
  \bibfield  {author} {\bibinfo {author} {\bibfnamefont {J.}~\bibnamefont
  {Vidal}}, \bibinfo {author} {\bibfnamefont {G.}~\bibnamefont {Palacios}}, \
  and\ \bibinfo {author} {\bibfnamefont {C.}~\bibnamefont {Aslangul}},\
  }\bibfield  {title} {\enquote {\bibinfo {title} {Entanglement dynamics in the
  {Lipkin-Meshkov-Glick} model},}\ }\href {\doibase 10.1103/PhysRevA.70.062304}
  {\bibfield  {journal} {\bibinfo  {journal} {Phys. Rev. A}\ }\textbf {\bibinfo
  {volume} {70}},\ \bibinfo {pages} {062304} (\bibinfo {year}
  {2004})}\BibitemShut {NoStop}%
\bibitem [{\citenamefont {Wilms}\ \emph {et~al.}(2012)\citenamefont {Wilms},
  \citenamefont {Vidal}, \citenamefont {Verstraete},\ and\ \citenamefont
  {Dusuel}}]{wilms_finite-temperature_2012}%
  \BibitemOpen
  \bibfield  {author} {\bibinfo {author} {\bibfnamefont {J.}~\bibnamefont
  {Wilms}}, \bibinfo {author} {\bibfnamefont {J.}~\bibnamefont {Vidal}},
  \bibinfo {author} {\bibfnamefont {F.}~\bibnamefont {Verstraete}}, \ and\
  \bibinfo {author} {\bibfnamefont {S.}~\bibnamefont {Dusuel}},\ }\bibfield
  {title} {{\selectlanguage {english}\enquote {\bibinfo {title}
  {Finite-temperature mutual information in a simple phase transition},}\
  }}\href {\doibase 10.1088/1742-5468/2012/01/P01023} {\bibfield  {journal}
  {\bibinfo  {journal} {J. Stat. Mech. Theor. Exp.}\ }\textbf {\bibinfo
  {volume} {2012}},\ \bibinfo {pages} {P01023} (\bibinfo {year}
  {2012})}\BibitemShut {NoStop}%
\bibitem [{\citenamefont {Pichler}\ \emph
  {et~al.}(2013{\natexlab{a}})\citenamefont {Pichler}, \citenamefont {Bonnes},
  \citenamefont {Daley}, \citenamefont {L\"auchli},\ and\ \citenamefont
  {Zoller}}]{pichler_thermal_2013}%
  \BibitemOpen
  \bibfield  {author} {\bibinfo {author} {\bibfnamefont {H.}~\bibnamefont
  {Pichler}}, \bibinfo {author} {\bibfnamefont {L.}~\bibnamefont {Bonnes}},
  \bibinfo {author} {\bibfnamefont {A.~J.}\ \bibnamefont {Daley}}, \bibinfo
  {author} {\bibfnamefont {A.~M.}\ \bibnamefont {L\"auchli}}, \ and\ \bibinfo
  {author} {\bibfnamefont {P.}~\bibnamefont {Zoller}},\ }\bibfield  {title}
  {{\selectlanguage {english}\enquote {\bibinfo {title} {Thermal versus
  entanglement entropy: a measurement protocol for fermionic atoms with a
  quantum gas microscope},}\ }}\href
  {http://iopscience.iop.org/1367-2630/15/6/063003} {\bibfield  {journal}
  {\bibinfo  {journal} {New Journal of Physics}\ }\textbf {\bibinfo {volume}
  {15}},\ \bibinfo {pages} {063003} (\bibinfo {year}
  {2013}{\natexlab{a}})}\BibitemShut {NoStop}%
\bibitem [{\citenamefont {Dicke}(1954)}]{dicke_coherence_1954}%
  \BibitemOpen
  \bibfield  {author} {\bibinfo {author} {\bibfnamefont {R.~H.}\ \bibnamefont
  {Dicke}},\ }\bibfield  {title} {\enquote {\bibinfo {title} {Coherence in
  spontaneous radiation processes},}\ }\href {\doibase 10.1103/PhysRev.93.99}
  {\bibfield  {journal} {\bibinfo  {journal} {Phys. Rev.}\ }\textbf {\bibinfo
  {volume} {93}},\ \bibinfo {pages} {99--110} (\bibinfo {year}
  {1954})}\BibitemShut {NoStop}%
\bibitem [{\citenamefont {Pfeuty}(1970)}]{pfeuty_one-dimensional_1970}%
  \BibitemOpen
  \bibfield  {author} {\bibinfo {author} {\bibfnamefont {P.}~\bibnamefont
  {Pfeuty}},\ }\bibfield  {title} {\enquote {\bibinfo {title} {The
  one-dimensional {Ising} model with a transverse field},}\ }\href {\doibase
  10.1016/0003-4916(70)90270-8} {\bibfield  {journal} {\bibinfo  {journal}
  {Ann. Phys.}\ }\textbf {\bibinfo {volume} {57}},\ \bibinfo {pages} {79--90}
  (\bibinfo {year} {1970})}\BibitemShut {NoStop}%
\bibitem [{\citenamefont {Garc{\'i}a-Ripoll}(2006)}]{garcia-ripoll_time_2006}%
  \BibitemOpen
  \bibfield  {author} {\bibinfo {author} {\bibfnamefont {J.~J.}\ \bibnamefont
  {Garc{\'i}a-Ripoll}},\ }\bibfield  {title} {\enquote {\bibinfo {title} {Time
  evolution of matrix product states},}\ }\href {\doibase
  10.1088/1367-2630/8/12/305} {\bibfield  {journal} {\bibinfo  {journal} {New
  J. Phys.}\ }\textbf {\bibinfo {volume} {8}},\ \bibinfo {pages} {305--305}
  (\bibinfo {year} {2006})}\BibitemShut {NoStop}%
\bibitem [{\citenamefont {Latorre}\ \emph {et~al.}(2005)\citenamefont
  {Latorre}, \citenamefont {Orus}, \citenamefont {Rico},\ and\ \citenamefont
  {Vidal}}]{latorre_entanglement_2005}%
  \BibitemOpen
  \bibfield  {author} {\bibinfo {author} {\bibfnamefont {J.~I.}\ \bibnamefont
  {Latorre}}, \bibinfo {author} {\bibfnamefont {R.}~\bibnamefont {Orus}},
  \bibinfo {author} {\bibfnamefont {E.}~\bibnamefont {Rico}}, \ and\ \bibinfo
  {author} {\bibfnamefont {J.}~\bibnamefont {Vidal}},\ }\bibfield  {title}
  {\enquote {\bibinfo {title} {Entanglement entropy in the
  {Lipkin-Meshkov-Glick} model},}\ }\href {\doibase 10.1103/PhysRevA.71.064101}
  {\bibfield  {journal} {\bibinfo  {journal} {Phys. Rev. A}\ }\textbf {\bibinfo
  {volume} {71}},\ \bibinfo {pages} {064101} (\bibinfo {year}
  {2005})}\BibitemShut {NoStop}%
\bibitem [{\citenamefont {Dutta}\ and\ \citenamefont
  {Bhattacharjee}(2001)}]{dutta_phase_2001}%
  \BibitemOpen
  \bibfield  {author} {\bibinfo {author} {\bibfnamefont {A.}~\bibnamefont
  {Dutta}}\ and\ \bibinfo {author} {\bibfnamefont {J.~K.}\ \bibnamefont
  {Bhattacharjee}},\ }\bibfield  {title} {\enquote {\bibinfo {title} {Phase
  transitions in the quantum {Ising} and rotor models with a long-range
  interaction},}\ }\href {\doibase 10.1103/PhysRevB.64.184106} {\bibfield
  {journal} {\bibinfo  {journal} {Phys. Rev. B}\ }\textbf {\bibinfo {volume}
  {64}},\ \bibinfo {pages} {184106} (\bibinfo {year} {2001})}\BibitemShut
  {NoStop}%
\bibitem [{\citenamefont {Koffel}\ \emph {et~al.}(2012)\citenamefont {Koffel},
  \citenamefont {Lewenstein},\ and\ \citenamefont
  {Tagliacozzo}}]{koffel_entanglement_2012}%
  \BibitemOpen
  \bibfield  {author} {\bibinfo {author} {\bibfnamefont {T.}~\bibnamefont
  {Koffel}}, \bibinfo {author} {\bibfnamefont {M.}~\bibnamefont {Lewenstein}},
  \ and\ \bibinfo {author} {\bibfnamefont {L.}~\bibnamefont {Tagliacozzo}},\
  }\bibfield  {title} {\enquote {\bibinfo {title} {Entanglement entropy for the
  long-range {Ising} chain in a transverse field},}\ }\href {\doibase
  10.1103/PhysRevLett.109.267203} {\bibfield  {journal} {\bibinfo  {journal}
  {Phys. Rev. Lett.}\ }\textbf {\bibinfo {volume} {109}},\ \bibinfo {pages}
  {267203} (\bibinfo {year} {2012})}\BibitemShut {NoStop}%
\bibitem [{\citenamefont {Deng}\ \emph {et~al.}(2005)\citenamefont {Deng},
  \citenamefont {Porras},\ and\ \citenamefont {Cirac}}]{deng_effective_2005}%
  \BibitemOpen
  \bibfield  {author} {\bibinfo {author} {\bibfnamefont {X.-L.}\ \bibnamefont
  {Deng}}, \bibinfo {author} {\bibfnamefont {D.}~\bibnamefont {Porras}}, \ and\
  \bibinfo {author} {\bibfnamefont {J.~I.}\ \bibnamefont {Cirac}},\ }\bibfield
  {title} {\enquote {\bibinfo {title} {Effective spin quantum phases in systems
  of trapped ions},}\ }\href {\doibase 10.1103/PhysRevA.72.063407} {\bibfield
  {journal} {\bibinfo  {journal} {Phys. Rev. A}\ }\textbf {\bibinfo {volume}
  {72}},\ \bibinfo {pages} {063407} (\bibinfo {year} {2005})}\BibitemShut
  {NoStop}%
\bibitem [{\citenamefont {Gardiner}(2004)}]{gardiner_handbook_2004}%
  \BibitemOpen
  \bibfield  {author} {\bibinfo {author} {\bibfnamefont {C.}~\bibnamefont
  {Gardiner}},\ }\href@noop {} {\emph {\bibinfo {title} {Handbook of Stochastic
  Methods}}},\ \bibinfo {edition} {3rd}\ ed.\ (\bibinfo  {publisher}
  {Springer},\ \bibinfo {address} {Berlin},\ \bibinfo {year}
  {2004})\BibitemShut {NoStop}%
\bibitem [{\citenamefont {Pichler}\ \emph
  {et~al.}(2013{\natexlab{b}})\citenamefont {Pichler}, \citenamefont
  {Schachenmayer}, \citenamefont {Daley},\ and\ \citenamefont
  {Zoller}}]{pichler_heating_2013}%
  \BibitemOpen
  \bibfield  {author} {\bibinfo {author} {\bibfnamefont {H.}~\bibnamefont
  {Pichler}}, \bibinfo {author} {\bibfnamefont {J.}~\bibnamefont
  {Schachenmayer}}, \bibinfo {author} {\bibfnamefont {A.~J.}\ \bibnamefont
  {Daley}}, \ and\ \bibinfo {author} {\bibfnamefont {P.}~\bibnamefont
  {Zoller}},\ }\bibfield  {title} {\enquote {\bibinfo {title} {Heating dynamics
  of bosonic atoms in a noisy optical lattice},}\ }\href {\doibase
  10.1103/PhysRevA.87.033606} {\bibfield  {journal} {\bibinfo  {journal} {Phys.
  Rev. A}\ }\textbf {\bibinfo {volume} {87}},\ \bibinfo {pages} {033606}
  (\bibinfo {year} {2013}{\natexlab{b}})}\BibitemShut {NoStop}%
\bibitem [{\citenamefont {H\"affner}\ \emph {et~al.}(2005)\citenamefont
  {H\"affner}, \citenamefont {H\"ansel}, \citenamefont {Roos}, \citenamefont
  {Benhelm}, \citenamefont {{Chek-al-kar}}, \citenamefont {Chwalla},
  \citenamefont {K\"orber}, \citenamefont {Rapol}, \citenamefont {Riebe},
  \citenamefont {Schmidt}, \citenamefont {Becher}, \citenamefont {G\"uhne},
  \citenamefont {D\"ur},\ and\ \citenamefont {Blatt}}]{haffner_scalable_2005}%
  \BibitemOpen
  \bibfield  {author} {\bibinfo {author} {\bibfnamefont {H.}~\bibnamefont
  {H\"affner}}, \bibinfo {author} {\bibfnamefont {W.}~\bibnamefont {H\"ansel}},
  \bibinfo {author} {\bibfnamefont {C.~F.}\ \bibnamefont {Roos}}, \bibinfo
  {author} {\bibfnamefont {J.}~\bibnamefont {Benhelm}}, \bibinfo {author}
  {\bibfnamefont {D.}~\bibnamefont {{Chek-al-kar}}}, \bibinfo {author}
  {\bibfnamefont {M.}~\bibnamefont {Chwalla}}, \bibinfo {author} {\bibfnamefont
  {T.}~\bibnamefont {K\"orber}}, \bibinfo {author} {\bibfnamefont {U.~D.}\
  \bibnamefont {Rapol}}, \bibinfo {author} {\bibfnamefont {M.}~\bibnamefont
  {Riebe}}, \bibinfo {author} {\bibfnamefont {P.~O.}\ \bibnamefont {Schmidt}},
  \bibinfo {author} {\bibfnamefont {C.}~\bibnamefont {Becher}}, \bibinfo
  {author} {\bibfnamefont {O.}~\bibnamefont {G\"uhne}}, \bibinfo {author}
  {\bibfnamefont {W.}~\bibnamefont {D\"ur}}, \ and\ \bibinfo {author}
  {\bibfnamefont {R.}~\bibnamefont {Blatt}},\ }\bibfield  {title}
  {{\selectlanguage {english}\enquote {\bibinfo {title} {Scalable multiparticle
  entanglement of trapped ions},}\ }}\href {\doibase 10.1038/nature04279}
  {\bibfield  {journal} {\bibinfo  {journal} {Nature}\ }\textbf {\bibinfo
  {volume} {438}},\ \bibinfo {pages} {643--646} (\bibinfo {year}
  {2005})}\BibitemShut {NoStop}%
\bibitem [{\citenamefont {Horodecki}\ and\ \citenamefont
  {Ekert}(2002)}]{horodecki_method_2002}%
  \BibitemOpen
  \bibfield  {author} {\bibinfo {author} {\bibfnamefont {P.}~\bibnamefont
  {Horodecki}}\ and\ \bibinfo {author} {\bibfnamefont {A.}~\bibnamefont
  {Ekert}},\ }\bibfield  {title} {\enquote {\bibinfo {title} {Method for direct
  detection of quantum entanglement},}\ }\href {\doibase
  10.1103/PhysRevLett.89.127902} {\bibfield  {journal} {\bibinfo  {journal}
  {Phys. Rev. Lett.}\ }\textbf {\bibinfo {volume} {89}},\ \bibinfo {pages}
  {127902} (\bibinfo {year} {2002})}\BibitemShut {NoStop}%
\bibitem [{\citenamefont {Cardy}(2011)}]{cardy_measuring_2011}%
  \BibitemOpen
  \bibfield  {author} {\bibinfo {author} {\bibfnamefont {J.}~\bibnamefont
  {Cardy}},\ }\bibfield  {title} {\enquote {\bibinfo {title} {Measuring
  entanglement using quantum quenches},}\ }\href {\doibase
  10.1103/PhysRevLett.106.150404} {\bibfield  {journal} {\bibinfo  {journal}
  {Phys. Rev. Lett.}\ }\textbf {\bibinfo {volume} {106}},\ \bibinfo {pages}
  {150404} (\bibinfo {year} {2011})}\BibitemShut {NoStop}%
\bibitem [{\citenamefont {Abanin}\ and\ \citenamefont
  {Demler}(2012)}]{abanin_measuring_2012}%
  \BibitemOpen
  \bibfield  {author} {\bibinfo {author} {\bibfnamefont {D.~A.}\ \bibnamefont
  {Abanin}}\ and\ \bibinfo {author} {\bibfnamefont {E.}~\bibnamefont
  {Demler}},\ }\bibfield  {title} {\enquote {\bibinfo {title} {Measuring
  entanglement entropy of a generic many-body system with a quantum switch},}\
  }\href {\doibase 10.1103/PhysRevLett.109.020504} {\bibfield  {journal}
  {\bibinfo  {journal} {Phys. Rev. Lett.}\ }\textbf {\bibinfo {volume} {109}},\
  \bibinfo {pages} {020504} (\bibinfo {year} {2012})}\BibitemShut {NoStop}%
\bibitem [{\citenamefont {Daley}\ \emph {et~al.}(2012)\citenamefont {Daley},
  \citenamefont {Pichler}, \citenamefont {Schachenmayer},\ and\ \citenamefont
  {Zoller}}]{daley_measuring_2012}%
  \BibitemOpen
  \bibfield  {author} {\bibinfo {author} {\bibfnamefont {A.~J.}\ \bibnamefont
  {Daley}}, \bibinfo {author} {\bibfnamefont {H.}~\bibnamefont {Pichler}},
  \bibinfo {author} {\bibfnamefont {J.}~\bibnamefont {Schachenmayer}}, \ and\
  \bibinfo {author} {\bibfnamefont {P.}~\bibnamefont {Zoller}},\ }\bibfield
  {title} {\enquote {\bibinfo {title} {Measuring entanglement growth in quench
  dynamics of bosons in an optical lattice},}\ }\href {\doibase
  10.1103/PhysRevLett.109.020505} {\bibfield  {journal} {\bibinfo  {journal}
  {Phys. Rev. Lett.}\ }\textbf {\bibinfo {volume} {109}},\ \bibinfo {pages}
  {020505} (\bibinfo {year} {2012})}\BibitemShut {NoStop}%
\bibitem [{\citenamefont {Schachenmayer}\ \emph {et~al.}(2010)\citenamefont
  {Schachenmayer}, \citenamefont {Lesanovsky}, \citenamefont {Micheli},\ and\
  \citenamefont {Daley}}]{schachenmayer_dynamical_2010}%
  \BibitemOpen
  \bibfield  {author} {\bibinfo {author} {\bibfnamefont {J.}~\bibnamefont
  {Schachenmayer}}, \bibinfo {author} {\bibfnamefont {I.}~\bibnamefont
  {Lesanovsky}}, \bibinfo {author} {\bibfnamefont {A.}~\bibnamefont {Micheli}},
  \ and\ \bibinfo {author} {\bibfnamefont {A.~J.}\ \bibnamefont {Daley}},\
  }\bibfield  {title} {\enquote {\bibinfo {title} {Dynamical crystal creation
  with polar molecules or rydberg atoms in optical lattices},}\ }\href
  {\doibase 10.1088/1367-2630/12/10/103044} {\bibfield  {journal} {\bibinfo
  {journal} {New J. Phys.}\ }\textbf {\bibinfo {volume} {12}},\ \bibinfo
  {pages} {103044} (\bibinfo {year} {2010})}\BibitemShut {NoStop}%
\bibitem [{\citenamefont {Gorshkov}\ \emph {et~al.}(2011)\citenamefont
  {Gorshkov}, \citenamefont {Manmana}, \citenamefont {Chen}, \citenamefont
  {Ye}, \citenamefont {Demler}, \citenamefont {Lukin},\ and\ \citenamefont
  {Rey}}]{gorshkov_tunable_2011}%
  \BibitemOpen
  \bibfield  {author} {\bibinfo {author} {\bibfnamefont {A.~V.}\ \bibnamefont
  {Gorshkov}}, \bibinfo {author} {\bibfnamefont {S.~R.}\ \bibnamefont
  {Manmana}}, \bibinfo {author} {\bibfnamefont {G.}~\bibnamefont {Chen}},
  \bibinfo {author} {\bibfnamefont {J.}~\bibnamefont {Ye}}, \bibinfo {author}
  {\bibfnamefont {E.}~\bibnamefont {Demler}}, \bibinfo {author} {\bibfnamefont
  {M.~D.}\ \bibnamefont {Lukin}}, \ and\ \bibinfo {author} {\bibfnamefont
  {A.~M.}\ \bibnamefont {Rey}},\ }\bibfield  {title} {\enquote {\bibinfo
  {title} {Tunable superfluidity and quantum magnetism with ultracold polar
  molecules},}\ }\href {\doibase 10.1103/PhysRevLett.107.115301} {\bibfield
  {journal} {\bibinfo  {journal} {Phys. Rev. Lett.}\ }\textbf {\bibinfo
  {volume} {107}},\ \bibinfo {pages} {115301} (\bibinfo {year}
  {2011})}\BibitemShut {NoStop}%
\bibitem [{\citenamefont {Chotia}\ \emph {et~al.}(2012)\citenamefont {Chotia},
  \citenamefont {Neyenhuis}, \citenamefont {Moses}, \citenamefont {Yan},
  \citenamefont {Covey}, \citenamefont {Foss-Feig}, \citenamefont {Rey},
  \citenamefont {Jin},\ and\ \citenamefont {Ye}}]{chotia_long-lived_2012}%
  \BibitemOpen
  \bibfield  {author} {\bibinfo {author} {\bibfnamefont {A.}~\bibnamefont
  {Chotia}}, \bibinfo {author} {\bibfnamefont {B.}~\bibnamefont {Neyenhuis}},
  \bibinfo {author} {\bibfnamefont {S.~A.}\ \bibnamefont {Moses}}, \bibinfo
  {author} {\bibfnamefont {B.}~\bibnamefont {Yan}}, \bibinfo {author}
  {\bibfnamefont {J.~P.}\ \bibnamefont {Covey}}, \bibinfo {author}
  {\bibfnamefont {M.}~\bibnamefont {Foss-Feig}}, \bibinfo {author}
  {\bibfnamefont {A.~M.}\ \bibnamefont {Rey}}, \bibinfo {author} {\bibfnamefont
  {D.~S.}\ \bibnamefont {Jin}}, \ and\ \bibinfo {author} {\bibfnamefont
  {J.}~\bibnamefont {Ye}},\ }\bibfield  {title} {\enquote {\bibinfo {title}
  {Long-lived dipolar molecules and {Feshbach} molecules in a {3D} optical
  lattice},}\ }\href {\doibase 10.1103/PhysRevLett.108.080405} {\bibfield
  {journal} {\bibinfo  {journal} {Phys. Rev. Lett.}\ }\textbf {\bibinfo
  {volume} {108}},\ \bibinfo {pages} {080405} (\bibinfo {year}
  {2012})}\BibitemShut {NoStop}%
\bibitem [{\citenamefont {Lesanovsky}(2011)}]{lesanovsky_many-body_2011}%
  \BibitemOpen
  \bibfield  {author} {\bibinfo {author} {\bibfnamefont {I.}~\bibnamefont
  {Lesanovsky}},\ }\bibfield  {title} {\enquote {\bibinfo {title} {Many-body
  spin interactions and the ground state of a dense {Rydberg} lattice gas},}\
  }\href {\doibase 10.1103/PhysRevLett.106.025301} {\bibfield  {journal}
  {\bibinfo  {journal} {Phys. Rev. Lett.}\ }\textbf {\bibinfo {volume} {106}},\
  \bibinfo {pages} {025301} (\bibinfo {year} {2011})}\BibitemShut {NoStop}%
\bibitem [{\citenamefont {Weimer}\ \emph {et~al.}(2010)\citenamefont {Weimer},
  \citenamefont {M\"uller}, \citenamefont {Lesanovsky}, \citenamefont
  {Zoller},\ and\ \citenamefont {B\"uchler}}]{weimer_rydberg_2010}%
  \BibitemOpen
  \bibfield  {author} {\bibinfo {author} {\bibfnamefont {H.}~\bibnamefont
  {Weimer}}, \bibinfo {author} {\bibfnamefont {M.}~\bibnamefont {M\"uller}},
  \bibinfo {author} {\bibfnamefont {I.}~\bibnamefont {Lesanovsky}}, \bibinfo
  {author} {\bibfnamefont {P.}~\bibnamefont {Zoller}}, \ and\ \bibinfo {author}
  {\bibfnamefont {H.~P.}\ \bibnamefont {B\"uchler}},\ }\bibfield  {title}
  {{\selectlanguage {english}\enquote {\bibinfo {title} {A {Rydberg} quantum
  simulator},}\ }}\href {\doibase 10.1038/nphys1614} {\bibfield  {journal}
  {\bibinfo  {journal} {Nat. Phys.}\ }\textbf {\bibinfo {volume} {6}},\
  \bibinfo {pages} {382--388} (\bibinfo {year} {2010})}\BibitemShut {NoStop}%
\bibitem [{\citenamefont {Viteau}\ \emph {et~al.}(2011)\citenamefont {Viteau},
  \citenamefont {Bason}, \citenamefont {Radogostowicz}, \citenamefont
  {Malossi}, \citenamefont {Ciampini}, \citenamefont {Morsch},\ and\
  \citenamefont {Arimondo}}]{viteau_rydberg_2011}%
  \BibitemOpen
  \bibfield  {author} {\bibinfo {author} {\bibfnamefont {M.}~\bibnamefont
  {Viteau}}, \bibinfo {author} {\bibfnamefont {M.~G.}\ \bibnamefont {Bason}},
  \bibinfo {author} {\bibfnamefont {J.}~\bibnamefont {Radogostowicz}}, \bibinfo
  {author} {\bibfnamefont {N.}~\bibnamefont {Malossi}}, \bibinfo {author}
  {\bibfnamefont {D.}~\bibnamefont {Ciampini}}, \bibinfo {author}
  {\bibfnamefont {O.}~\bibnamefont {Morsch}}, \ and\ \bibinfo {author}
  {\bibfnamefont {E.}~\bibnamefont {Arimondo}},\ }\bibfield  {title} {\enquote
  {\bibinfo {title} {{Rydberg} excitations in {Bose-Einstein} condensates in
  quasi-one-dimensional potentials and optical lattices},}\ }\href {\doibase
  10.1103/PhysRevLett.107.060402} {\bibfield  {journal} {\bibinfo  {journal}
  {Phys. Rev. Lett.}\ }\textbf {\bibinfo {volume} {107}},\ \bibinfo {pages}
  {060402} (\bibinfo {year} {2011})}\BibitemShut {NoStop}%
\bibitem [{\citenamefont {Schau{\ss}}\ \emph {et~al.}(2012)\citenamefont
  {Schau{\ss}}, \citenamefont {Cheneau}, \citenamefont {Endres}, \citenamefont
  {Fukuhara}, \citenamefont {Hild}, \citenamefont {Omran}, \citenamefont
  {Pohl}, \citenamefont {Gross}, \citenamefont {Kuhr},\ and\ \citenamefont
  {Bloch}}]{schaus_observation_2012}%
  \BibitemOpen
  \bibfield  {author} {\bibinfo {author} {\bibfnamefont {P.}~\bibnamefont
  {Schau{\ss}}}, \bibinfo {author} {\bibfnamefont {M.}~\bibnamefont {Cheneau}},
  \bibinfo {author} {\bibfnamefont {M.}~\bibnamefont {Endres}}, \bibinfo
  {author} {\bibfnamefont {T.}~\bibnamefont {Fukuhara}}, \bibinfo {author}
  {\bibfnamefont {S.}~\bibnamefont {Hild}}, \bibinfo {author} {\bibfnamefont
  {A.}~\bibnamefont {Omran}}, \bibinfo {author} {\bibfnamefont
  {T.}~\bibnamefont {Pohl}}, \bibinfo {author} {\bibfnamefont {C.}~\bibnamefont
  {Gross}}, \bibinfo {author} {\bibfnamefont {S.}~\bibnamefont {Kuhr}}, \ and\
  \bibinfo {author} {\bibfnamefont {I.}~\bibnamefont {Bloch}},\ }\bibfield
  {title} {{\selectlanguage {english}\enquote {\bibinfo {title} {Observation of
  spatially ordered structures in a two-dimensional {Rydberg} gas},}\ }}\href
  {\doibase 10.1038/nature11596} {\bibfield  {journal} {\bibinfo  {journal}
  {Nature}\ }\textbf {\bibinfo {volume} {491}},\ \bibinfo {pages} {87--91}
  (\bibinfo {year} {2012})}\BibitemShut {NoStop}%
\bibitem [{\citenamefont {Hauke}\ and\ \citenamefont
  {Tagliacozzo}(2013)}]{hauke_spread_2013}%
  \BibitemOpen
  \bibfield  {author} {\bibinfo {author} {\bibfnamefont {P.}~\bibnamefont
  {Hauke}}\ and\ \bibinfo {author} {\bibfnamefont {L.}~\bibnamefont
  {Tagliacozzo}},\ }\bibfield  {title} {\enquote {\bibinfo {title} {Spread of
  correlations in long-range interacting systems},}\ }\href@noop {} {\bibfield
  {journal} {\bibinfo  {journal} {{arXiv:1304.7725}}\ } (\bibinfo {year}
  {2013})}\BibitemShut {NoStop}%
\bibitem [{\citenamefont {Amico}\ \emph {et~al.}(2008)\citenamefont {Amico},
  \citenamefont {Fazio}, \citenamefont {Osterloh},\ and\ \citenamefont
  {Vedral}}]{amico_entanglement_2008}%
  \BibitemOpen
  \bibfield  {author} {\bibinfo {author} {\bibfnamefont {L.}~\bibnamefont
  {Amico}}, \bibinfo {author} {\bibfnamefont {R.}~\bibnamefont {Fazio}},
  \bibinfo {author} {\bibfnamefont {A.}~\bibnamefont {Osterloh}}, \ and\
  \bibinfo {author} {\bibfnamefont {V.}~\bibnamefont {Vedral}},\ }\bibfield
  {title} {\enquote {\bibinfo {title} {Entanglement in many-body systems},}\
  }\href {\doibase 10.1103/RevModPhys.80.517} {\bibfield  {journal} {\bibinfo
  {journal} {Rev. Mod. Phys.}\ }\textbf {\bibinfo {volume} {80}},\ \bibinfo
  {pages} {517--576} (\bibinfo {year} {2008})}\BibitemShut {NoStop}%
\bibitem [{\citenamefont {Eisert}\ \emph {et~al.}(2010)\citenamefont {Eisert},
  \citenamefont {Cramer},\ and\ \citenamefont
  {Plenio}}]{eisert_colloquium:_2010}%
  \BibitemOpen
  \bibfield  {author} {\bibinfo {author} {\bibfnamefont {J.}~\bibnamefont
  {Eisert}}, \bibinfo {author} {\bibfnamefont {M.}~\bibnamefont {Cramer}}, \
  and\ \bibinfo {author} {\bibfnamefont {M.~B.}\ \bibnamefont {Plenio}},\
  }\bibfield  {title} {\enquote {\bibinfo {title} {Colloquium: Area laws for
  the entanglement entropy},}\ }\href {\doibase 10.1103/RevModPhys.82.277}
  {\bibfield  {journal} {\bibinfo  {journal} {Rev. Mod. Phys.}\ }\textbf
  {\bibinfo {volume} {82}},\ \bibinfo {pages} {277--306} (\bibinfo {year}
  {2010})}\BibitemShut {NoStop}%
\bibitem [{\citenamefont {Sidje}(1998)}]{sidje_expokit:_1998}%
  \BibitemOpen
  \bibfield  {author} {\bibinfo {author} {\bibfnamefont {R.~B.}\ \bibnamefont
  {Sidje}},\ }\bibfield  {title} {\enquote {\bibinfo {title} {Expokit: a
  software package for computing matrix exponentials},}\ }\href {\doibase
  10.1145/285861.285868} {\bibfield  {journal} {\bibinfo  {journal} {{ACM}
  Trans. Math. Softw.}\ }\textbf {\bibinfo {volume} {24}},\ \bibinfo {pages}
  {130Ð156} (\bibinfo {year} {1998})}\BibitemShut {NoStop}%
\end{thebibliography}%

\end{document}